\documentclass[manuscript]{aastex}
\usepackage{amsmath}
\usepackage{lscape}
\usepackage{comment}
\usepackage{grffile}
\usepackage{multirow}

\newcommand{\ocs}{$J=19-18$}
\newcommand{\MN}{CH$_3$OH}
\newcommand{\mna}{$5_{1,5}-4_{1,4}$; A$^+$}

\newcommand{\mnsourcea}{11$_{0, 11}-10_{1, 10}$; A$^{+}$}
\newcommand{\MF}{HCOOCH$_3$}
\newcommand{\mfa}{$20_{3,17}-19_{3,16}$; A}

\newcommand{\TFA}{H$_2$CS}
\newcommand{\tfaa}{$7_{0,7}-6_{0,6}$} 
\newcommand{\tfab}{$7_{2,5}-6_{2,4}$} 
 
\newcommand{\tfada}{$7_{4,4}-6_{4,3}$}
\newcommand{\tfadb}{$7_{4,3}-6_{4,2}$}
\newcommand{\tfad}{\tfada, \tfadb} 
\newcommand{\tfae}{$7_{4}-6_{4}$} 

\newcommand{\sio}{$J=6-5$}

\newcommand{\iras}{IRAS 16293--2422}
\newcommand{\ire}{infalling-rotating envelope}
\newcommand{\cb}{centrifugal barrier} 
\newcommand{\iPC}{inverse P-Cygni profile}
\newcommand{\desys}{disk/envelope system}

\newcommand{\Msun}{$M_\odot$}

\newcommand{\rcb}{$r_{\rm CB}$}

\newcommand{\vlsr}{$v_{\rm LSR}$}
\newcommand{\vsys}{$v_{\rm sys}$}

\newcommand{\inv}{$^{-1}$}
\newcommand{\kmps}{km s\inv}
\newcommand{\fdegr}{.\!\!\degr}
\newcommand{\hydro}{H$_2$}
\newcommand{\nhydro}{$n$(\hydro)}

\newcommand{\cmcubic}{cm$^{-3}$}

\newcommand{\PAenv}{110\degr}
\newcommand{\PAoutflow}{200\degr}

\newcommand{\parM}{0.4 \Msun}
\newcommand{\parRcb}{40}
\newcommand{\parRcbau}{\parRcb\ au}

\newcommand{\parI}{5\degr}
\newcommand{\incRem}{0\degr\ for a face-on configuration}
\newcommand{\parTau}{50 au}
\newcommand{\parRTFAau}{300 au}
\newcommand{\parRfall}{35}

\newcommand{\parLW}{0.5 \kmps}
\newcommand{\parVsys}{2.9 \kmps}
\newcommand{\aboutVsys}{$\sim 3$ \kmps}

\newcommand{\offsetCB}{60}
\newcommand{\offsetEnv}{120}

\newcommand{\Ka}{$K_{\rm a}$}

\newcommand{\iffigure}{\iftrue}

\renewcommand{\bf}{}

\shorttitle{IRE in I16293B}
\shortauthors{Oya et al.}

\title{Chemical and Physical Picture of IRAS 16293--2422 Source B at a Sub-arcsecond Scale Studied with ALMA}
\author{Yoko Oya\altaffilmark{1}, Kana Moriwaki\altaffilmark{1}, Shusuke Onishi\altaffilmark{1}, Nami Sakai\altaffilmark{2}, Ana L\'{o}pez--Sepulcre\altaffilmark{3, 4, 5}, \\
C\'{e}cile Favre\altaffilmark{6}, Yoshimasa Watanabe\altaffilmark{1, 7, 8}, Cecilia Ceccarelli\altaffilmark{4, 5}, Bertrand Lefloch\altaffilmark{4, 5}, \\and Satoshi Yamamoto\altaffilmark{1}} 

\email{oya@taurus.phys.s.u-tokyo.ac.jp}
\altaffiltext{1}{Department of Physics, The University of Tokyo, 7-3-1, Hongo, Bunkyo-ku, Tokyo 113-0033, Japan}
\altaffiltext{2}{The Institute of Physical and Chemical Research (RIKEN), Wako, Saitama 351-0198, Japan}
\altaffiltext{3}{Institut de Radioastronomie Millim\'etrique (IRAM), 38406, Saint Martin d'H\'eres, France}
\altaffiltext{4}{Universite Grenoble Alpes, IPAG, F-38000 Grenoble, France}
\altaffiltext{5}{CNRS, IPAG, F-38000 Grenoble, France}
\altaffiltext{6}{INAF, Osservatorio Astrofisico di Arcetri, Largo E. Fermi 5, 50125 Firenze, Italy}
\altaffiltext{7}{Division of Physics, Faculty of Pure and Applied Sciences, University of Tsukuba,  Tsukuba, Ibaraki 305-8571, Japan}
\altaffiltext{8}{Tomonaga Center for the History of the Universe, Faculty of Pure and Applied Sciences, University of Tsukuba,  Tsukuba, Ibaraki 305-8571, Japan}

\begin{abstract}
We have analyzed the OCS, H$_2$CS, CH$_3$OH, and HCOOCH$_3$ data observed 
toward the low-mass protostar IRAS 16293--2422 Source B at a sub-arcsecond resolution with ALMA. 
A clear chemical differentiation is seen in their distributions; 
OCS and H$_2$CS are extended with a slight rotation signature, 
while CH$_3$OH and HCOOCH$_3$ are concentrated near the protostar. 
Such a chemical change in the vicinity of the protostar is similar to the companion (Source A) case. 
The extended component is interpreted by the infalling-rotating envelope model with a nearly face-on configuration. 
The radius of the centrifugal barrier of the infalling-rotating envelope is roughly evaluated to be ($30-50$) au. 
The observed lines show the inverse P-Cygni profile, 
indicating the infall motion with in a few 10 au from the protostar. 
The nearly pole-on geometry of the outflow lobes is inferred from the SiO distribution, 
and thus, the infalling and outflowing motions should coexist along the line-of-sight to the protostar. 
This implies that the infalling gas is localized near the protostar and the current launching points of the outflow have an offset from the protostar. 
A possible mechanism for this configuration is discussed. 
\end{abstract}


\begin{document}
\section{Introduction} \label{sec:16293}
\iras\ is a well-studied Class 0 protostellar source in Ophiuchus ($d = 120$ pc) \citep{KnudeHog1998}. 
This source is known to be a binary, consisting of Source A and Source B, whose apparent separation is 
5\arcsec\ \citep[$\sim$600 au; e.g.][]{Wootten_binary, Mundy_binary, Bottinelli_Mass}. 
Molecular gas distribution and dynamics around the binary components have extensively been studied 
at a high angular resolution with millimeter- and submillimeter-wave interferometers 
\citep[e.g.][]{Mundy_binary, Bottinelli_Mass, Kuan_COMs, Pineda_ALMA, Zapata_infall, Favre_SMA, Jorgensen_SMA, Jorgensen_Cycle1}. 
This source is also famous as a prototypical hot corino source, 
which is characterized by rich complex organic molecules (COMs) in the vicinity of the protostars 
\citep[e.g.][]{Schoier_hotcore, Cazaux_hotcore, Bottinelli_Mass, Kuan_COMs, Pineda_ALMA, Jorgensen_sugar, Jorgensen_Cycle1, Coutens_PILS}. 

Recently, ALMA Cycle 1 archival data toward this source was analyzed 
to investigate the kinematic structure of the \ire\ associated with Source A by \citet{Oya_16293A}. 
The authors found that the kinematic structure of the \ire\ is successfully explained by a simple ballistic model \citep{Oya_15398}. 
Based on this study, the radius of the \cb\ of the \ire\ was evaluated to be $\sim$50 au. 
Since the \cb\ in the \ire\ is also identified in other low-mass protostellar sources, 
L1527, TMC-1A, L483, BHB07-11, and HH212 
\citep{Sakai_1527nature, Sakai_1527apjl, Sakai_TMC1A, Sakai_1527_highres, Oya_1527, Oya_483, Alves_BHB07-11, Lee_HH212}, 
it seems to be a common occurrence in low-mass star formation. 

In addition, a salient result of the above study \citep[i.e.][]{Oya_16293A} is that a drastic chemical change was found around the \cb; 
the OCS (carbonyl sulphide) and \TFA\ (thioformaldehyde) lines trace the \ire\ outside the \cb, 
while the COM lines, such as \MN\ (methanol) and \MF\ (methyl formate), mainly highlight the \cb. 
The \TFA\ lines also trace the high velocity component inside the \cb, which is likely a circumstellar disk component. 
Such enhancement of the COM emission around the \cb\ would be an essential part of the hot corino chemistry in this source. 
Recently, a similar distribution of COMs around the \cb\ is also reported for the low-mass protostellar source HH212 by \citet{Lee_HH212}. 
Thus, the \cb\ seems to stand for not only the physical transition zone from the \ire\ to the disk, 
but also the transition zone of the chemical composition. 
Although the species tracing each part is different, 
chemical changes around \cb s are reported for other sources, 
L1527, TMC-1A, and L483 \citep{Sakai_1527apjl, Sakai_TMC1A, Oya_483}. 	
Conversely, chemical diagnostics can be a powerful method to investigate the physical structure of the gas around a protostar at a 100 au scale. 

In this study, we apply the chemical diagnostics to \iras\ Source B, which is the other component of the binary system. 
Source B is known to be rich in COMs as well as Source A 
\citep[e.g.][]{Bottinelli_Mass, Kuan_COMs, Jorgensen_SMA, Jorgensen_sugar, Jorgensen_Cycle1, Pineda_ALMA}. 
Its disk/envelope system is reported to have a nearly face-on geometry in contrast to the edge-on geometry of Source A 
\citep[e.g.][]{Pineda_ALMA, Zapata_infall, Oya_16293A}. 
For this reason, the molecular line emission shows a narrower linewidth toward Source B than toward Source A. 
Furthermore, an \iPC\ is reported toward Source B 
\citep[e.g.][]{Pineda_ALMA, Jorgensen_sugar}, 
which implies the existence of the infalling gas in front of the protostar along the line-of-sight. 
In this study, we analyze molecular distributions and the kinematic structure of this source at a sub-arcsecond resolution, 
and compare the results with those of Source A. 

\section{Observations} \label{sec:observation}
In this study, we used the ALMA Cycle 1 archival data of \iras\ in Band 6 (\#2012.1.00712.S), 
which covers the frequency ranges from 230 to 250 GHz and from 220 to 240 GHz. 
The details of the observations are reported by \citet{Jorgensen_Cycle1} and \citet{Oya_16293A}. 
Here, we briefly summarize important points. 
In the observations, (42--44) antennas were used, 
and the primary beam (half-power beam width) is (24--25)\arcsec. 
The largest recoverable scale is $\sim$13\arcsec\ at 240 GHz. 
The backend correlator was tuned to a resolution of 122 kHz, 
which corresponds to the velocity resolution of 0.15 km s$^{-1}$ at 240 GHz, 
and a bandwidth of 468.750 MHz. 
The data calibration was performed in the antenna-based manner and uncertainties are less than 10 \% \citep[ALMA Cycle 1 Technical Handbook; ][]{TechHB_Cyc1}. 

In addition, 
we also analyzed the ALMA Cycle 3 data, which were carried out on 5 March 2016 (\#2015.1.01060.S). 
41 antennas were used in these observations with the baseline length ranging from 17 to 636 m. 
The largest recoverable scale is $\sim$14\arcsec\ at 260 GHz. 
The phase center of these observations is ($\alpha_{2000}$, $\delta_{2000}$) = ($16^{\rm h} 32^{\rm m} 22\fs87$, $-24\degr 28\arcmin 36\farcs3$), 
and the primary beam is 25\arcsec. 
The total on-source time was 16.38 minutes with typical system temperature of (60--140) K. 
The backend correlator was tuned to a resolution of 122 kHz, which corresponds to the velocity resolution of 0.14 \kmps\ at 260 GHz, 
and a bandwidth of 58.6 MHz. 
J1625-2527 was used for the phase calibration. 
The bandpass calibration was done on the quasar J1427-4206, 
and the absolute flux density scale was derived from Titan. 
The data calibration was performed in the antenna-based manner, 
and uncertainties are less than 10 \% \citep[ALMA Cycle 3 Technical Handbook; ][]{TechHB_Cyc3}. 

Observed spectral lines are summarized in Table \ref{tb:lines}, 
including their rest frequencies, upper state energies, intrinsic line strengths, and synthesized beams. 
The 1.3 mm (236 GHz) continuum image was obtained by averaging the line-free channels of the two Cycle 1 data (230 and 240 GHz), 
whose total band width is 7.5 GHz. 
The line images were obtained by subtracting the continuum data directly from the visibilities. 
The Brigg's weighting with the robustness parameter of 0.5 was employed to obtain the images of the continuum and the spectral lines. 
Self-calibration using the continuum emission was applied to the ALMA Cycle 1 data (the continuum, OCS, \MN, \MF, and \TFA). 
On the other hand, it was not applied to the ALMA Cycle 3 data (SiO), 
because the continuum sensitivity is not enough for self-calibration processing 
due to a limited number of line-free channels. 

\section{Distribution} \label{sec:distribution}
\subsection{Overall Distributions} \label{sec:dist_all}
\begin{figure}
	\iffigure
	\epsscale{1.0}
	\includegraphics[viewport = 0 0 1000 500, scale = 0.7]{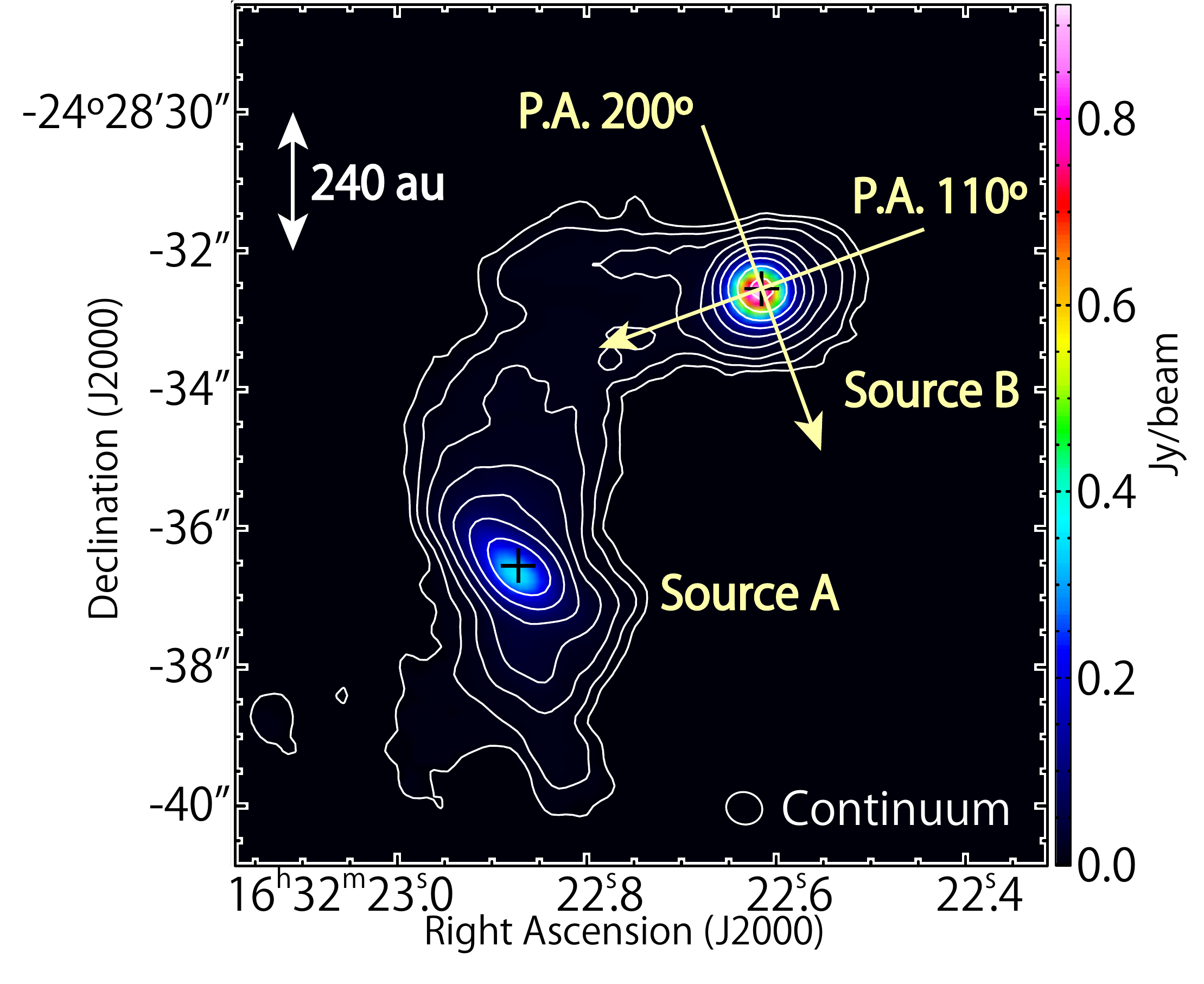}
	\fi
	\caption{Map of the continuum emission at the 1.3 mm (236 GHz) band. 
			The contour levels are 10, 20, 40, 80, 160, 320, 640, 1280, and 2560$\sigma$, where the rms noise level is 0.3 mJy beam\inv. 
			The intensity peak positions are shown by the black crosses. 
			The synthesized beam is depicted in the bottom-right corner. 
			PA of \PAenv\ and \PAoutflow\ indicate the disk/envelope direction and the direction perpendicular to it. 
			\label{fig:cont}}
\end{figure}

Figure \ref{fig:cont} shows the 1.3 mm (236 GHz) continuum map. 
The synthesized beam is $(0\farcs524 \times 0\farcs463)$ (PA $73\fdegr87$). 
There are two intensity peaks corresponding to the two components of the binary, Source A and Source B. 
A weak emission bridging between the two peaks can also be seen. 
These features are consistent with the previous report by \citet{Jorgensen_Cycle1}. 
While the distribution around Source A is slightly elongated along the NE-SW direction, 
that around Source B has an almost round shape. 
The disk/envelope systems of Source A and B are reported to have nearly edge-on and face-on configurations, 
respectively \citep[e.g.][]{Rodriguez_16293B, Chandler_infall}, 
to which the above distributions are consistent. 
The full-width half-maximum (FWHM) sizes, the peak flux densities, 
and the peak position for Source A and Source B are 
determined by a two-dimensional Gaussian fit to the image. 
The FWHM sizes 
deconvolved by the beam are evaluated to be 
$(1\farcs138 \pm 0\farcs007) \times (0\farcs583 \pm 0\farcs004)$ (PA $36\fdegr1 \pm 0\fdegr35$) and 
$(0\farcs429 \pm 0\farcs003) \times (0\farcs357 \pm 0\farcs003)$ (PA $139\fdegr0 \pm 1\fdegr6$) 
for Source A and Source B, respectively. 
The peak flux densities are $(314.0 \pm 1.4)$ and $(899.9 \pm 1.5)$ mJy/beam for Source A and Source B, respectively. 
The derived continuum peak positions are: 
($\alpha_{2000}$, $\delta_{2000}$) = ($16^{\rm h} 32^{\rm m} 22\fs8725 \pm 0\fs0001$, $-24\degr 28\arcmin 36\farcs536 \pm 0\farcs002$) and 
($\alpha_{2000}$, $\delta_{2000}$) = ($16^{\rm h} 32^{\rm m} 22\fs61531 \pm 0\fs00003$, $-24\degr 28\arcmin 32\farcs5467 \pm 0\farcs0004$), 
for Source A and Source B, respectively. 
Although they may be affected by the self-calibration, the effect seems to be negligible for a beam size of $\sim0\farcs5$. 
Indeed, the continuum peak position of Source B was derived to be 
($\alpha_{2000}$, $\delta_{2000}$) = ($16^{\rm h} 32^{\rm m} 22\fs6142 \pm 0\fs0001$, $-24\degr 28\arcmin 32\farcs524\pm 0\farcs002$) 
without self-calibration processing, 
and the difference from the position derived from the self-calibrated data is very small ($\sim0\farcs03$). 

\begin{landscape}
\begin{figure}
	\iffigure
	\epsscale{0.8}
	\includegraphics[viewport = -50 0 100 500, scale = 0.75]{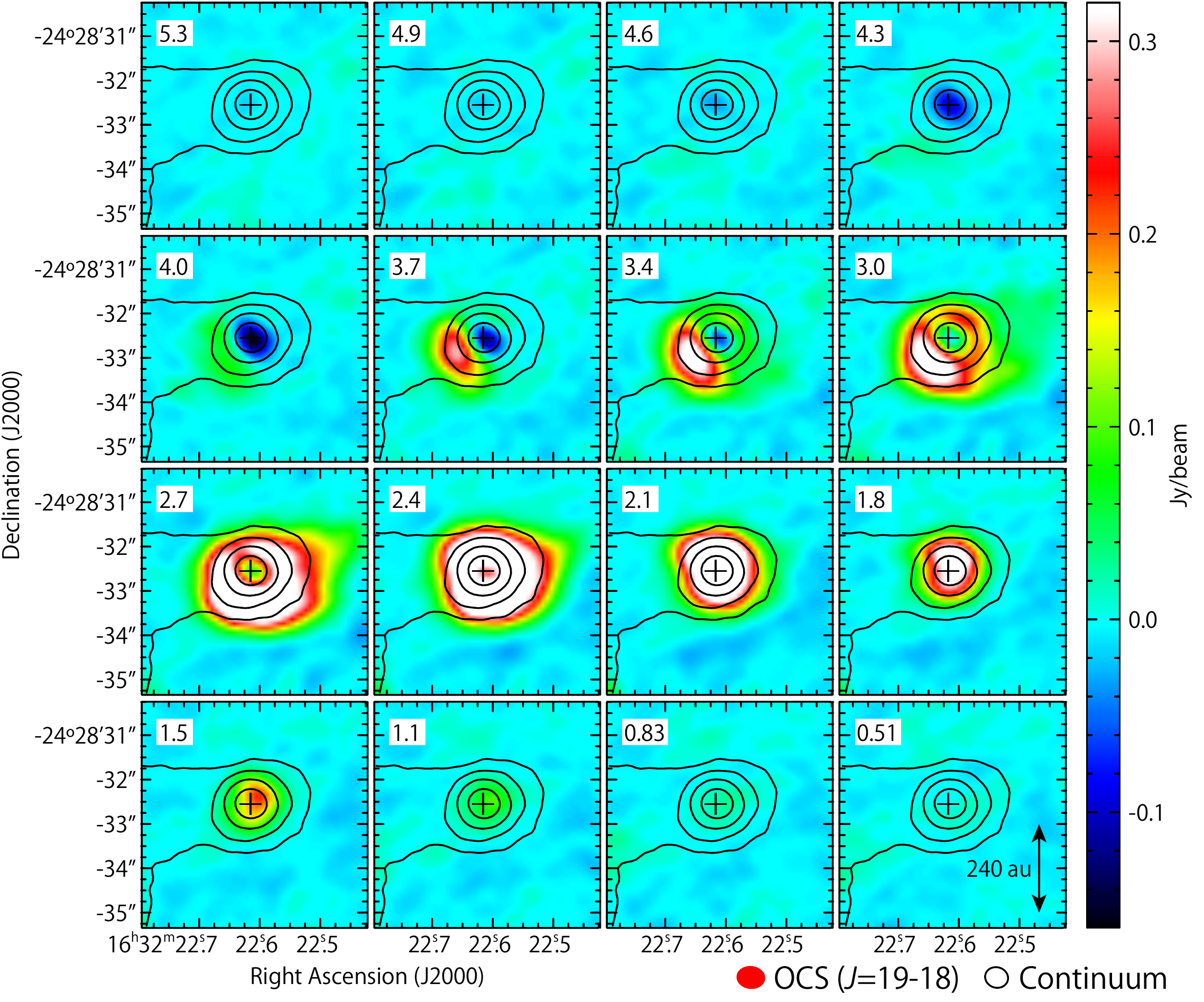} 
	\fi
	\caption{The velocity channel maps of the OCS (\ocs) line. 
			The \vlsr\ is shown in the left-upper corner of each panel, 
			where the systemic velocity is \aboutVsys. 
			The contours represent the continuum map (see Figure \ref{fig:cont}), 
			whose intensity peak position is shown by the black crosses. 
			The contour levels are 20, 80, 320, and 1280$\sigma$, where the rms noise level is 0.3 mJy beam\inv. 
			The synthesized beams are depicted at the bottom. 
			\label{fig:channelmap_OCS}}
\end{figure}
\end{landscape}

Figure \ref{fig:channelmap_OCS} shows the velocity channel maps of the OCS (\ocs) line. 
Absorption toward the continuum peak position can be seen in the channels with \vlsr\ ranging from 3.4 to 4.3 \kmps. 
Since the systemic velocity is around 3 \kmps\ \citep{Bottinelli_Mass}, this absorption feature is red-shifted. 
It is most naturally interpreted as the \iPC, as previously reported by \citet{Jorgensen_sugar}, \citet{Pineda_ALMA}, and \citet{Zapata_infall}. 
At the systemic velocity, the distribution is extended at a 3\arcsec\ ($\sim$400 au) scale in diameter around the continuum peak, 
although it would suffer from the resolving-out effect with the largest recoverable scale of 
$\sim$14\arcsec\ in this observation. 

\begin{landscape}
\begin{figure}
	\iffigure
	\epsscale{0.8}
	\includegraphics[viewport = -50 0 100 500, scale = 0.75]{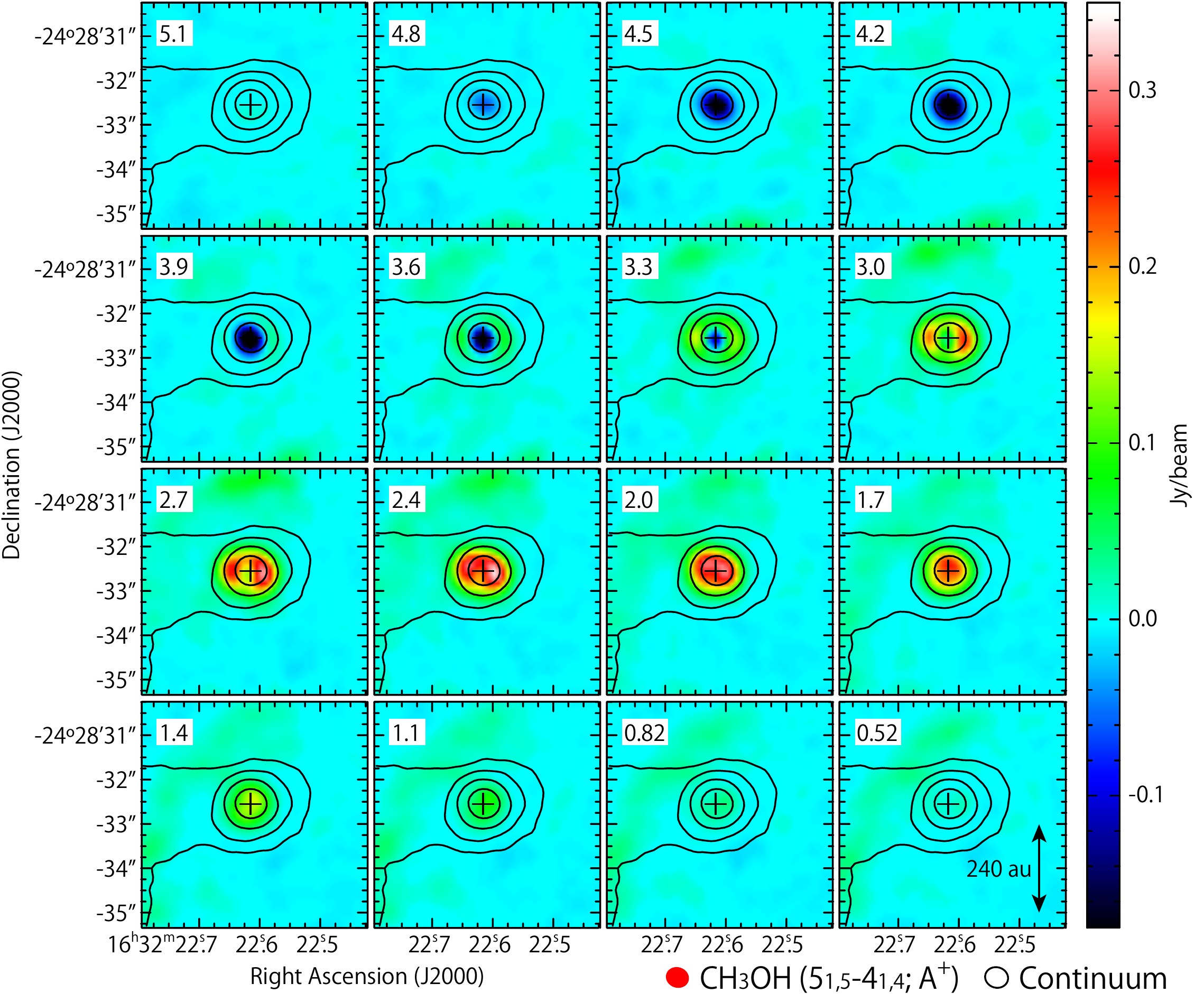} 
	\fi
	\caption{The velocity channel maps of the \MN\ (\mna) line. 
			The \vlsr\ is shown in the left-upper corner of each panel, 
			where the systemic velocity is \aboutVsys. 
			The contours represent the continuum map (see Figure \ref{fig:cont}), 
			whose intensity peak position is shown by the black crosses. 
			The contour levels are the same as those in Figure \ref{fig:channelmap_OCS}. 
			The synthesized beams are depicted at the bottom. 
			\label{fig:channelmap_MN}} 
\end{figure}
\end{landscape}

\begin{landscape}
\begin{figure}
	\iffigure
	\epsscale{0.8}
	\includegraphics[viewport = -50 0 100 500, scale = 0.75]{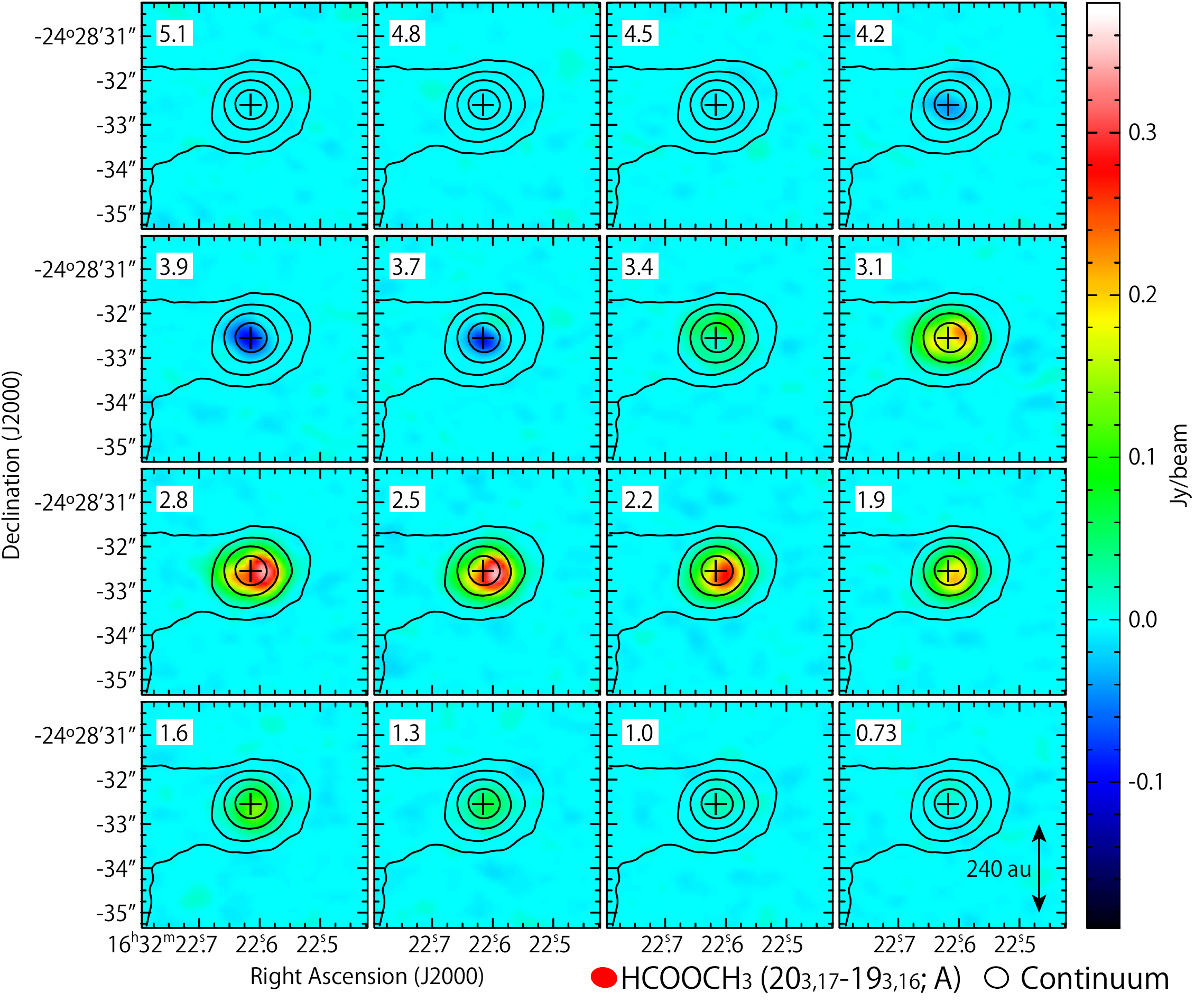} 
	\fi
	\caption{The velocity channel maps of the \MF\ (\mfa) line. 
			The \vlsr\ is shown in the left-upper corner of each panel, 
			where the systemic velocity is \aboutVsys. 
			The contours represent the continuum map (see Figure \ref{fig:cont}), 
			whose intensity peak position is shown by the black crosses. 
			The contour levels are the same as those in Figure \ref{fig:channelmap_OCS}. 
			The synthesized beams are depicted at the bottom. 
			\label{fig:channelmap_MF}} 
\end{figure}
\end{landscape}

\begin{landscape}
\begin{figure}
	\iffigure
	\epsscale{0.8}
	\includegraphics[viewport = -50 0 100 500, scale = 0.75]{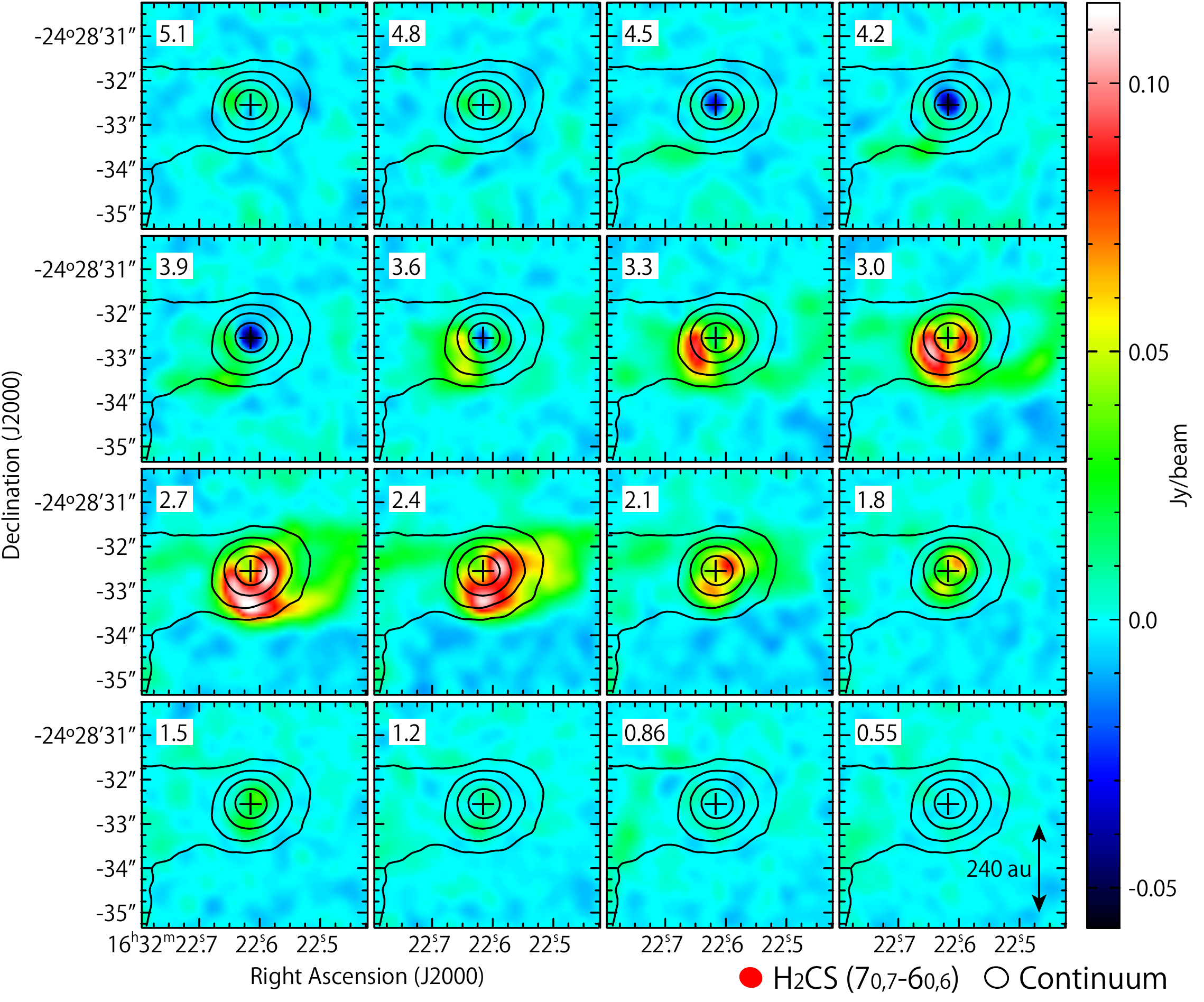} 
	\fi
	\caption{The velocity channel maps of the \TFA\ (\tfaa) line. 
			The \vlsr\ is shown in the left-upper corner of each panel, 
			where the systemic velocity is \aboutVsys. 
			The contours represent the continuum map (see Figure \ref{fig:cont}), 
			whose intensity peak position is shown by the black crosses. 
			The contour levels are the same as those in Figure \ref{fig:channelmap_OCS}. 
			The synthesized beams are depicted at the bottom. 
			\label{fig:channelmap_TFA}}
\end{figure}
\end{landscape}

Figures \ref{fig:channelmap_MN}, \ref{fig:channelmap_MF}, and \ref{fig:channelmap_TFA} 
show the velocity channel maps of the \MN\ (\mna), \MF\ (\mfa), and \TFA\ (\tfaa) lines, respectively. 
For all the three lines, the red-shifted components of the emission (\vlsr\ = 3--5 \kmps) 
show absorption features toward the continuum peak as in the OCS case. 
The distributions of \MN\ and \MF\ are more compact than that of OCS. 
Although the distribution of \TFA\ is also smaller than that of OCS around the protostar, 
it is larger than those of \MN\ and \MF. 
The \TFA\ emission is slightly extended toward the western side of the continuum peak. 

\begin{landscape}
\begin{figure}
	\iffigure
	\epsscale{1.0}
	\includegraphics[viewport = 50 0 100 1000, scale = 0.35]{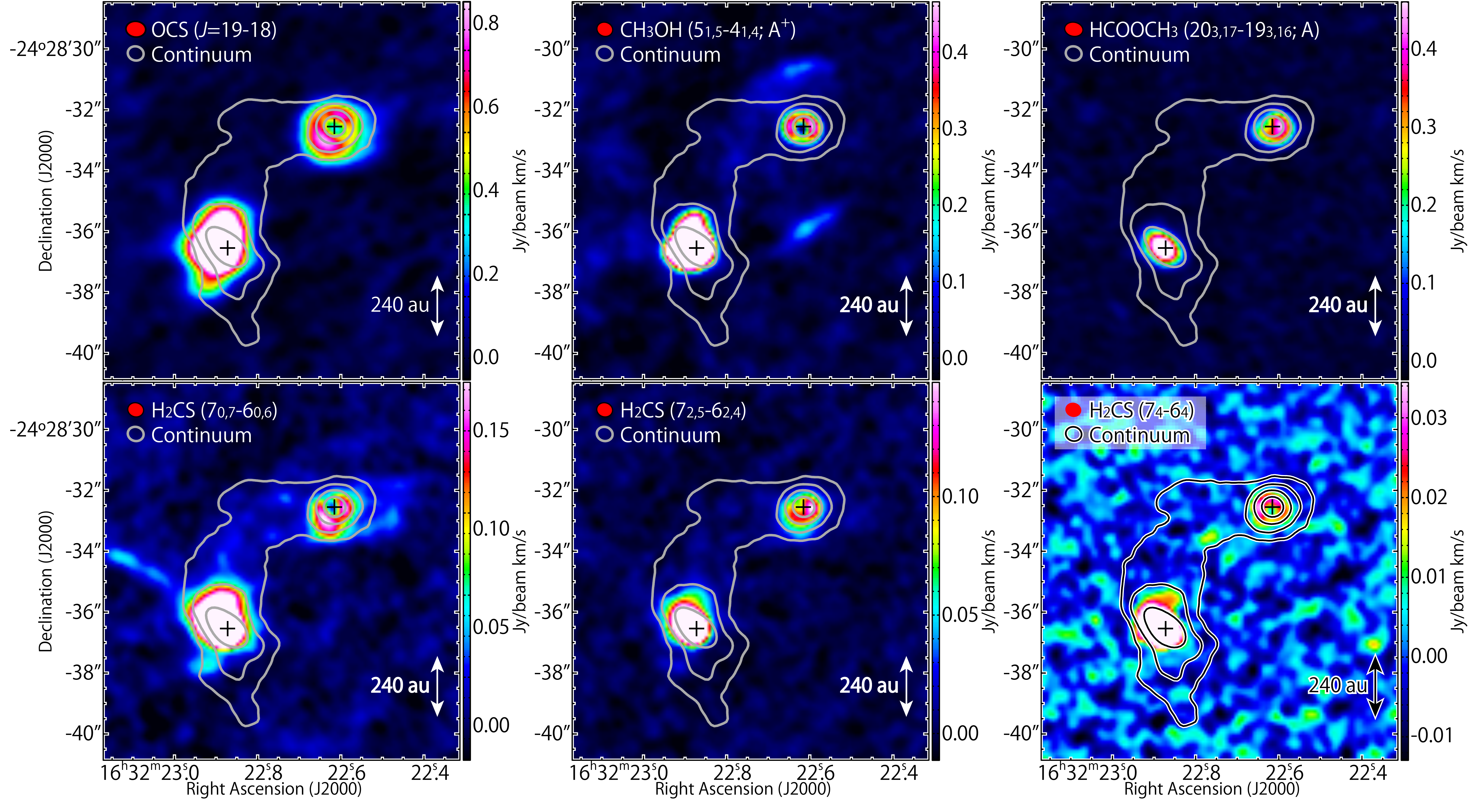}
	\fi
	\caption{The integrated intensity maps of the OCS (\ocs), \MN\ (\mna), \MF\ (\mfa), and \TFA\ (\tfaa; \tfab; \tfae) lines associated with Source B (color). 
			The contours represent the continuum map, 
			where the contour levels are the same as those in Figure \ref{fig:cont}. 
			The continuum peak positions are shown by the black crosses. 
			The synthesized beam are depicted in the top-left corners. 
			\label{fig:mom0}}
\end{figure}
\end{landscape}

The integrated intensity maps of 
the four molecular species are shown in Figure \ref{fig:mom0}. 
The intensity distributions, except for that of \MF, show a ring-like structure around the protostar; 
the molecular line intensities are weaker within a radius of 0\farcs25 ($\sim$30 au) toward the continuum peak position  
than the surrounding positions.  
This is due to the contribution of the \iPC\ toward the continuum peak position, as shown in the channel maps 
(Figures \ref{fig:channelmap_OCS}, \ref{fig:channelmap_MN}, and \ref{fig:channelmap_TFA}). 

The different sizes of the molecular distributions seen in the channel maps can also be confirmed in the integrated intensity maps. 
Indeed, 
the OCS distribution is clearly extended over 1\arcsec\ ($\sim$120 au) around the protostar. 
The \MN\ and \MF\ emission is concentrated around the protostar with a radius of 0\farcs6 ($\sim$70 au). 
It should be noted that 
a weak emission of \MN\ is seen 
at the angular distance of 2\arcsec\ and 3\arcsec\ in the northern and southern sides of the protostar, respectively. 
Although their origin is puzzling, these components may be related to the outflow 
originating from Source B (see Section \ref{sec:kinematics}), 
or may be enhanced by an interaction between the envelope gas and the outflow from Source A, 
as suggested from the CO ($J=6-5$) observations 
\citep{Kristensen_16293outflow}. 
Probably, extended components of the outflows would be resolved out in the present study. 
The \TFA\ (\tfaa) emission is slightly more extended than the \MN\ and \MF\ emission, 
but is not so extended as the OCS emission. 
The distributions of the higher excitation lines of \TFA\ (\tfab; \tfae) tend to be as compact as those of the \MN\ and \MF\ lines. 

\subsection{Chemical Differentiation} \label{sec:dist_diff}
\begin{figure}
	\iffigure
	\epsscale{1.0}
	\includegraphics[viewport = 0 0 100 200, scale = 0.53]{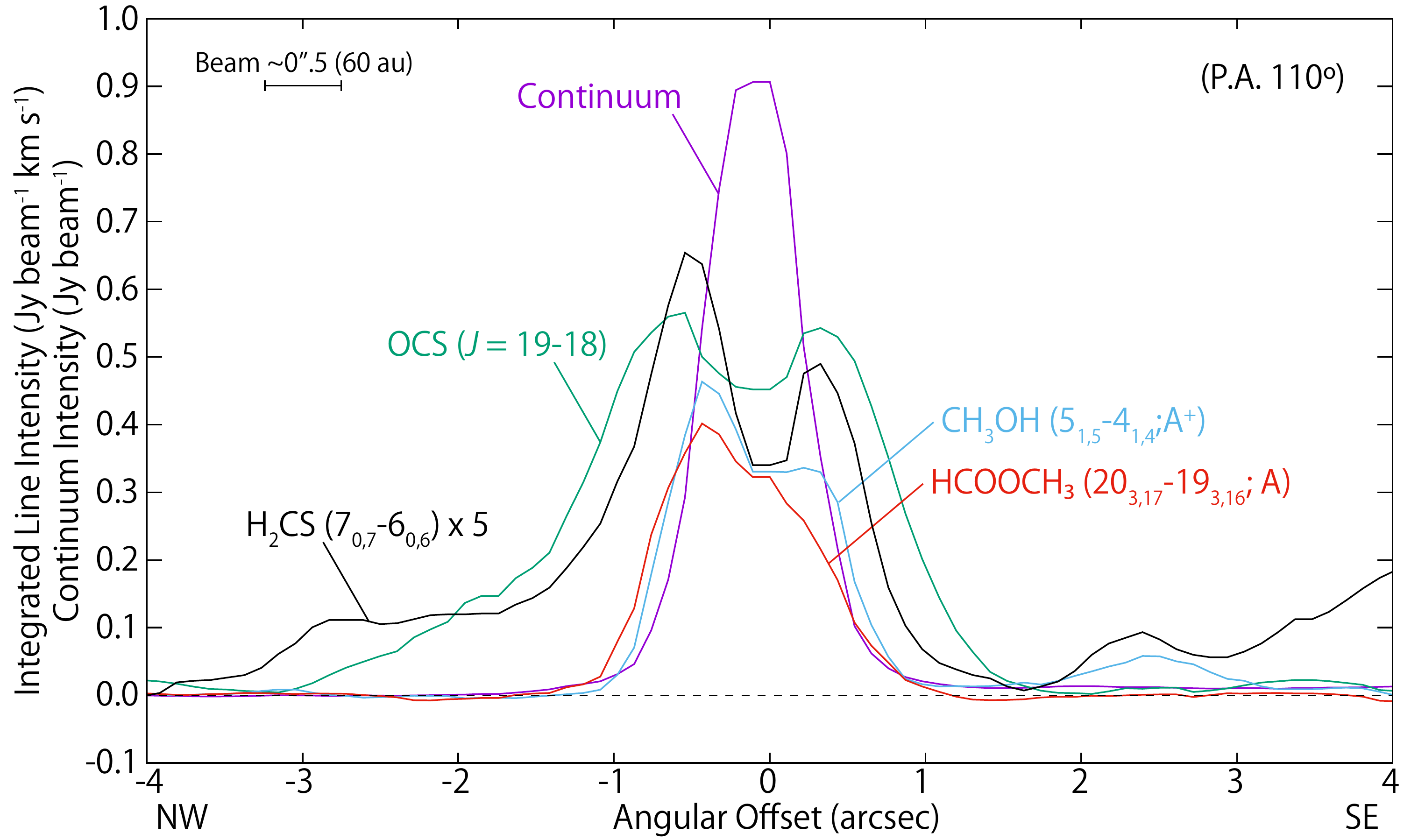}
	\fi
	\caption{The spatial profile of the integrated intensities of the observed molecular lines, 
			where the integrated velocity range is from 0.9 to 2.9 \kmps\ (\vsys\ = 2.9 \kmps). 
			The position axis is centered at the continuum peak position, 
			and its position angle is \PAenv\ (see Figures \ref{fig:cont} and \ref{fig:highV_OCS}). 
			This PA corresponds to the axis along which the \desys\ is extended (see Section \ref{sec:kin_rot}; PA \PAenv). 
			\label{fig:intprofile}}
\end{figure}

\begin{figure}
	\iffigure
	\epsscale{1.0}
	\includegraphics[viewport = 0 0 100 500, scale = 0.75]{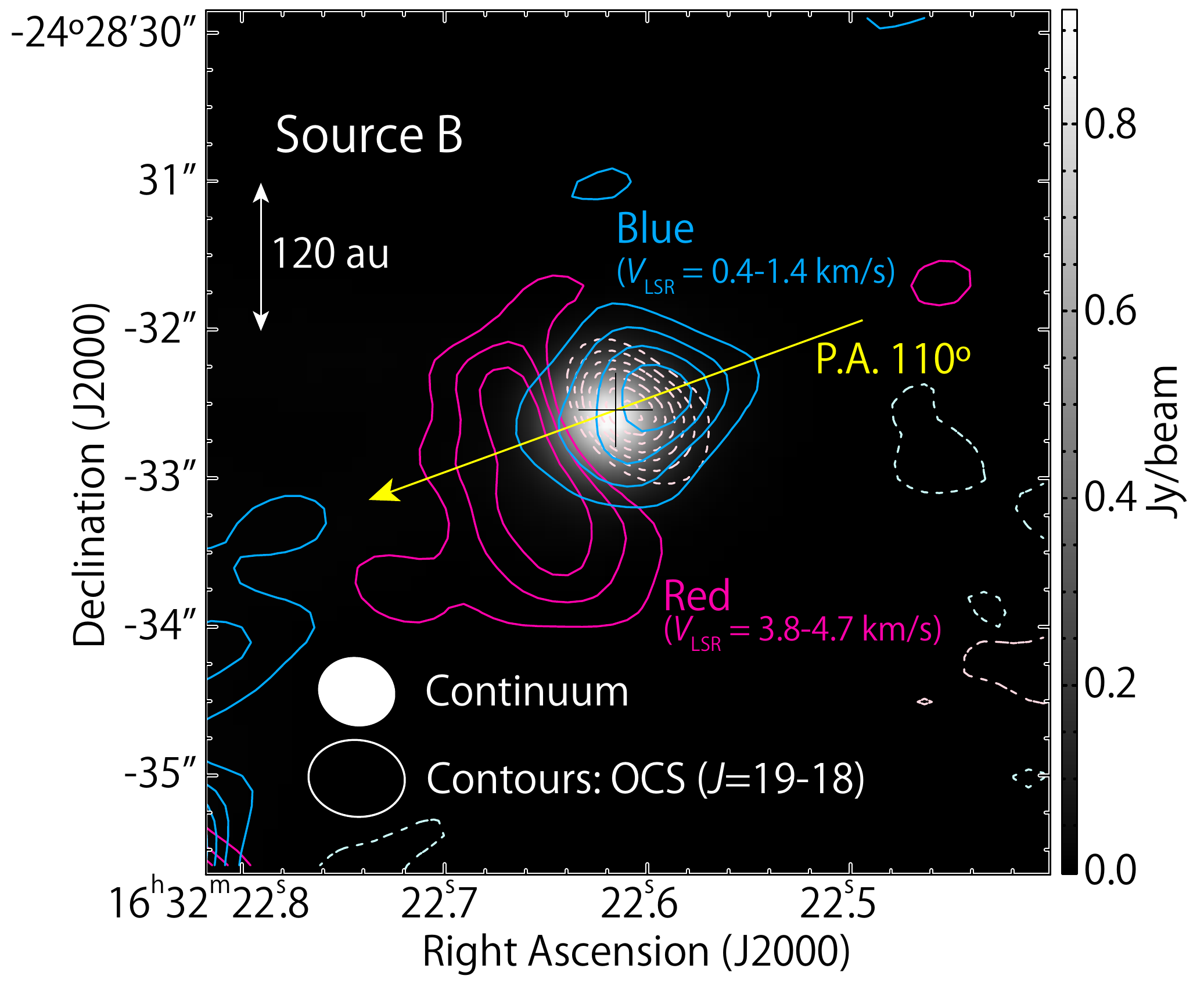}
	\fi
	\caption{The integrated intensity maps of high velocity components of the OCS (\ocs; contours) line. 
			The red contours represent the integrated intensity map of OCS with the velocity range from 3.8 to 4.7 \kmps, 
			while the blue contours with the velocity range from 0.4 to 1.4 \kmps. 
			The dashed contours represent the negative values. 
			The contour levels are every 3$\sigma$, where the rms noise level is 5 mJy beam\inv\ \kmps. 
			The gray scale map represent the continuum map, and the intensity peak position in Source B is shown by the black cross. 
			The synthesized beams are depicted in the bottom-left corner. 
			The PA of \PAenv\ shows the disk/envelope direction (see Section \ref{sec:kin_rot}). 
			\label{fig:highV_OCS}}
\end{figure}

The above results for OCS, \MN, \MF, and \TFA\ show small-scale chemical differentiation in the vicinity of the protostar in Source B. 
This differentiation is clearly confirmed by the spatial profiles of the integrated line intensities of these molecules, 
as shown in Figure \ref{fig:intprofile}. 
Only the blue-shifted components, 
which are in the velocity range from 0.9 to 2.9 \kmps, 
are integrated to prevent contamination of the \iPC. 
{\bf 
The position axis of the spatial profiles has the origin at the continuum peak, 
and is prepared along the line where the \desys\ is extended. 
Its position angle is \PAenv, as described in Section \ref{sec:kin_rot} (Figures \ref{fig:cont} and \ref{fig:highV_OCS}). 
}

In \iras\ Source B, 
it is well-known that the molecular line emission 
associated with the main molecular isotopologues 
is often optically thick \citep[e.g.,][]{Jorgensen_Cycle1}. 
In such a case, the line intensities do not reflect their actual abundances, 
especially toward the continuum peak. 
Thus, we need to consider the optical depth effect carefully. 
However, we confirm by a non-LTE excitation calculation
that the optical depth effect does not affect the difference of the emitting regions shown in Figure \ref{fig:intprofile}. 
Here, we used the RADEX code \citep{vanderTak_radex} 
assuming a kinetic temperature of 100 K and a \nhydro\ of $10^8$ \cmcubic. 
In this calculation, the optical depths of the OCS and \MN\ lines are roughly estimated to be about 0.2 and 0.01, respectively, 
at the angular offset of $-1$\arcsec\ (corresponding to $-120$ au) from the continuum peak in Figure \ref{fig:intprofile}, 
where their integrated intensities are 0.45 and 0.03 Jy beam\inv\ \kmps, respectively. 
The excitation temperatures are 100 and 97 K for the OCS and \MN\ lines, respectively, 
indicating that the lines are well thermalized. 

{\bf 
Since the excitation temperatures employed by \citet{Coutens_PILS} are between 100 and 300 K at 0\farcs5 from the protostar, 
the kinetic temperature would be higher than 70 K at 1\farcs0. 
Even in this condition, 
the optical depths of OCS and \MN\ are estimated to be lower than 0.3 according to the RADEX code. 
It should be noted that 
the above results for the optical depths do not change significantly for a \nhydro\ from $10^6$ to $10^9$ \cmcubic. 
}

Thus, the gradual change in the OCS intensity and the \MN\ intensity 
can be attributed to the difference in their respective distributions. 
In addition, 
the excitation effect cannot make the OCS distribution more extended 
in comparison with that of \MN\ and \MF, 
because the upper-state energy for the OCS line is comparable to that of the \MF\ line 
and is even higher than those of the \MN\ and \TFA\ lines (Table \ref{tb:lines}). 
Thus, the different distributions seem to mainly represent the change in the chemical composition of the gas. 

A similar chemical differentiation is also seen in Source A; 
the distributions of \MN\ and \MF\ are concentrated around the \cb\ with the radius of 50 au, 
while the OCS and \TFA\ distributions are more extended \citep{Oya_16293A}. 
In that context, 
\citet{Miura_2017} recently reported on the basis of numerical simulations 
that the enhancement of the COM lines around the \cb\ can be caused by 
the dust heating in accretion shocks in front of the \cb. 

\section{Kinematic Structure} \label{sec:kinematics}
\subsection{Rotation Feature} \label{sec:kin_rot}

The velocity widths (FWHM) of these line emissions toward Source B are as narrow as 3 \kmps. 
They are narrower than those toward Source A, which are typically wider than 5 \kmps. 
The narrow velocity widths likely originate from the nearly face-on geometry of the disk/envelope system of Source B, 
as mentioned in Section \ref{sec:16293}. 
Nevertheless, we can recognize small velocity gradient in the channel maps of OCS (Figure \ref{fig:channelmap_OCS}). 
This result indicates that the \desys\ is slightly inclined. 
At the blue-shifted velocity (\vlsr\ $\sim$1.5 \kmps), 
the OCS distribution shows a slight offset from the continuum peak position toward the northwestern (NW) direction, 
while at the red-shifted velocity (\vlsr\ $\sim$4 \kmps), 
it tends to have a slight offset toward the southeastern (SE) direction. 
This trend can be confirmed 
in the integrated intensity maps of the high velocity-shifted components (Figure \ref{fig:highV_OCS}); 
although the red-shifted component may be contaminated by the absorption toward the continuum peak, 
the blue-shifted component of the OCS line 
seems to be aligned on a straight line along the SE-NW direction with a position angle (PA) of about \PAenv. 
This velocity gradient suggests a rotation motion of the disk/envelope system, slightly inclined from the face-on geometry. 
In the below analysis, we use the position angle of \PAenv\ as the direction 
along which the mid-plane of the \desys\ is extended (hereafter `the disk/envelope direction'). 

\subsection{Observed Features} \label{sec:kin_obs}
\begin{figure}
	\iffigure
	\vspace*{-10pt}
	\epsscale{1.0}
	\includegraphics[viewport = 0 0 100 700, scale = 0.58]{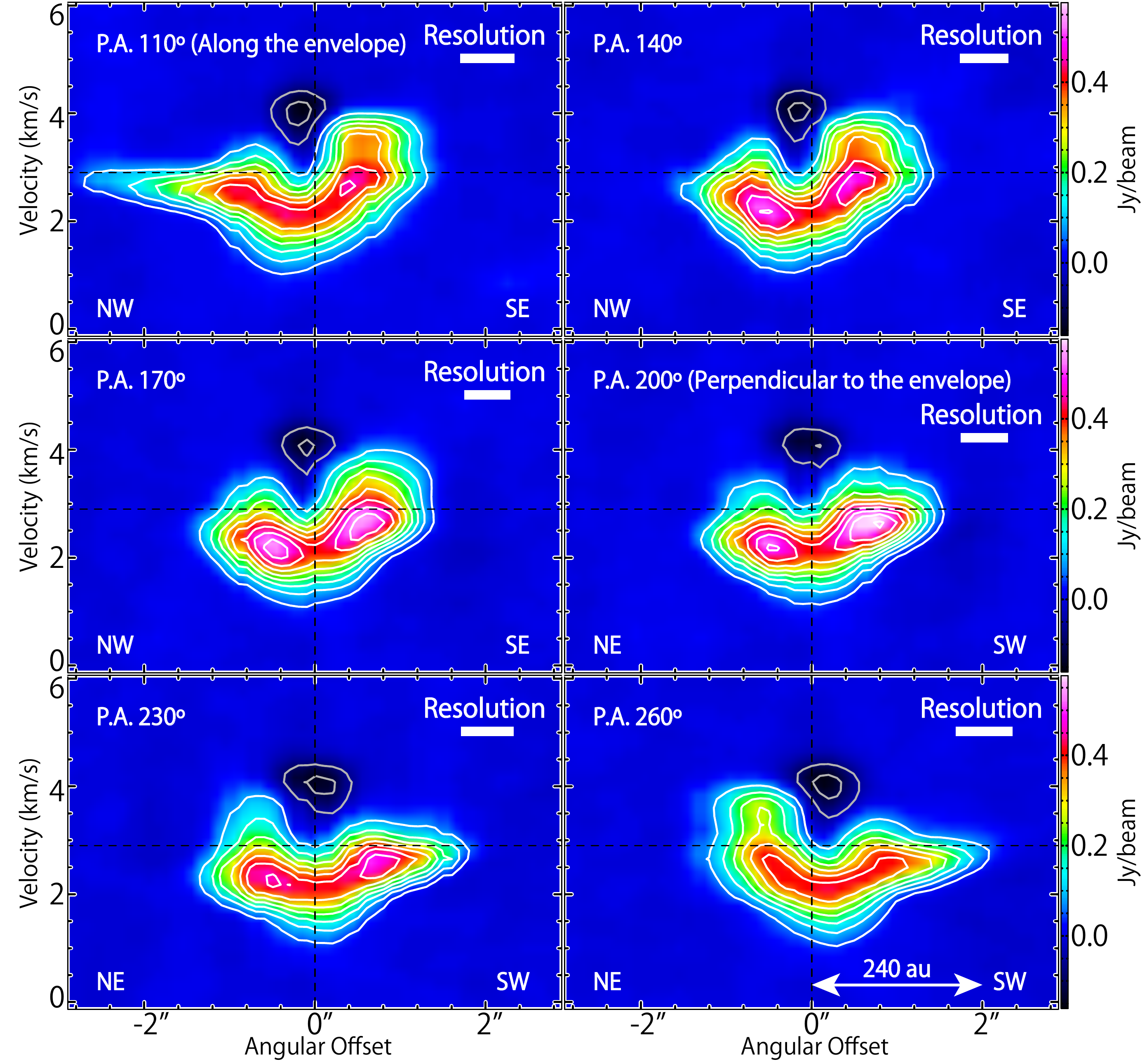}
	\fi
	\caption{The position-velocity diagrams of the OCS (\ocs) line. 
			The position axes are centered at the continuum peak position 
			and their position angles are every 30\degr\ from \PAenv. 
			The contour levels are from $-40\sigma$ and every 20$\sigma$, except for 0$\sigma$, 
			where the rms noise level is 3.2 mJy beam\inv. 
			The rectangle in each panel represents the spatial and velocity resolutions. 
			\label{fig:PV_OCS}}
\end{figure}

\begin{figure}
	\iffigure
	\vspace*{-10pt}
	\epsscale{1.0}
	\includegraphics[viewport = 0 0 100 700, scale = 0.58]{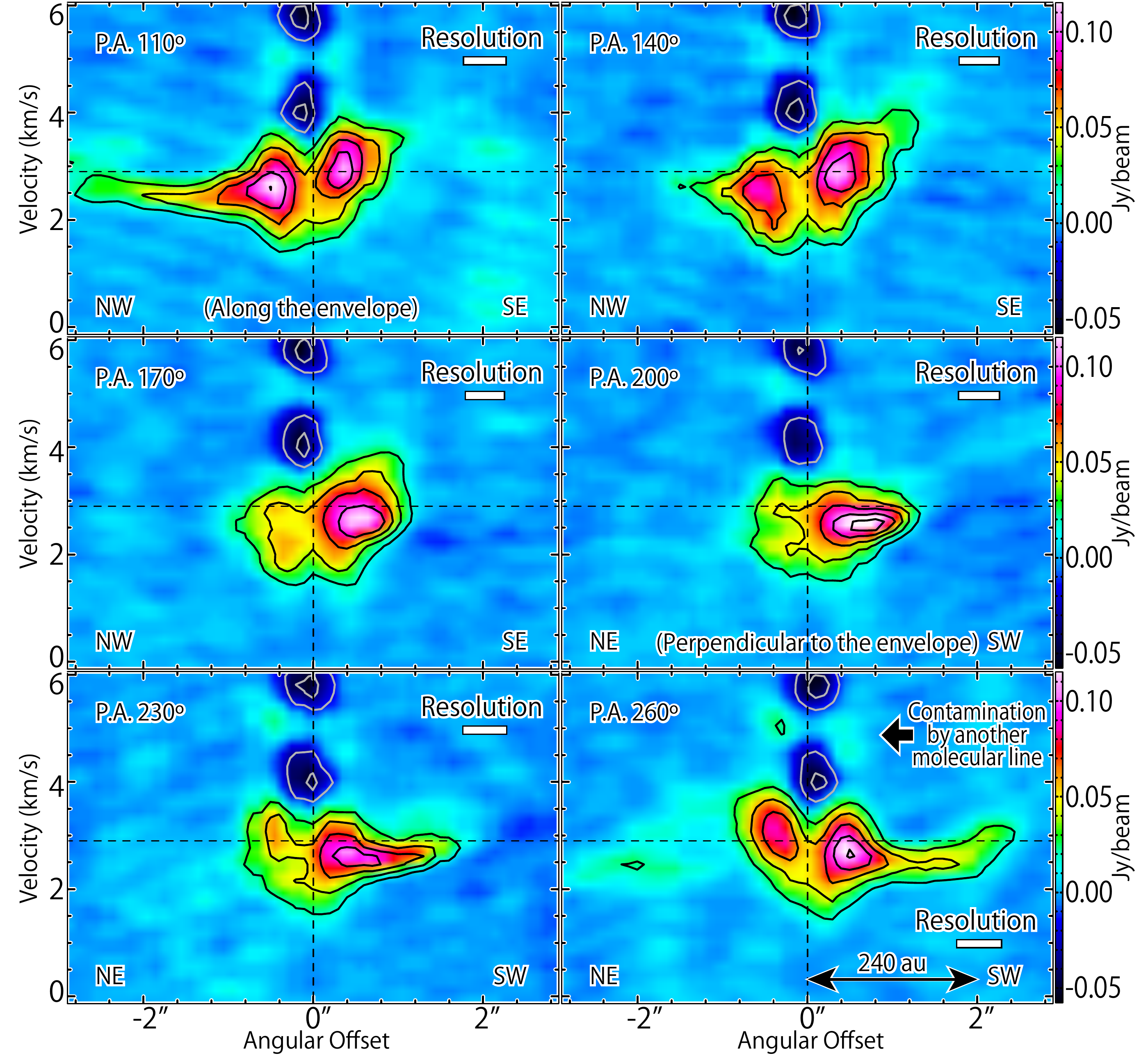}
	\fi
	\caption{The position-velocity diagrams of the \TFA\ (\tfaa) line. 
			The position axes are the same as those in Figure \ref{fig:PV_OCS}. 
			The contour levels are from -20$\sigma$ and every 10$\sigma$, except for 0$\sigma$, 
			where the rms noise level is 2.3 mJy beam\inv. 
			There is a contamination by another molecular line with a velocity offset of 2 \kmps. 
			The rectangle in each panel represents the spatial and velocity resolutions. 
			\label{fig:PV_TFA}}
\end{figure}

\begin{figure}
	\iffigure
	\vspace*{-10pt}
	\epsscale{1.0}
	\includegraphics[viewport = 0 0 100 700, scale = 0.58]{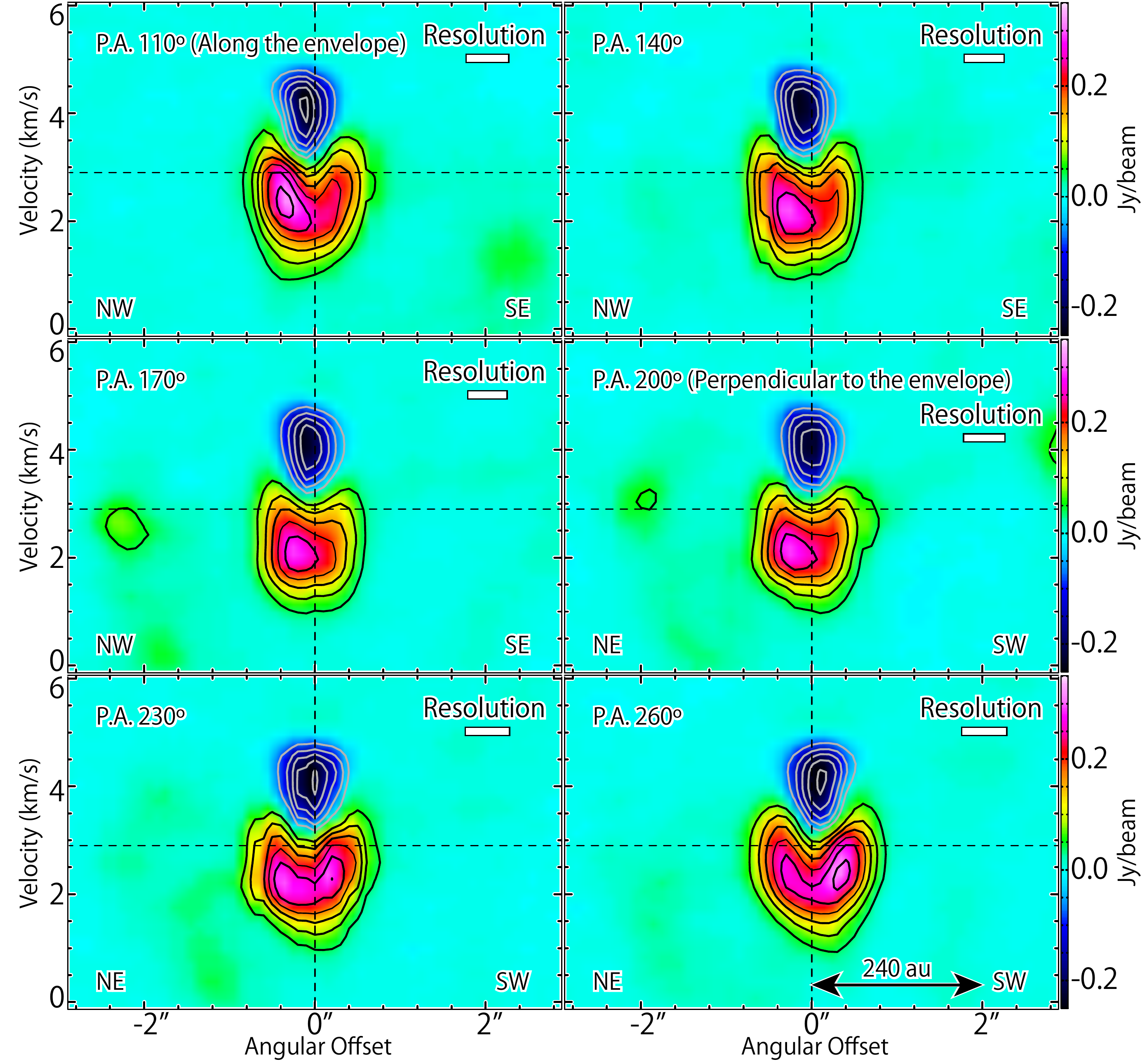}
	\fi
	\caption{The position-velocity diagrams of the \MN\ (\mna) line. 
			The position axes are the same as those in Figure \ref{fig:PV_OCS}. 
			The contour levels are from -100$\sigma$ and every 20$\sigma$, except for 0$\sigma$, 
			where the rms noise level is 2.5 mJy beam\inv. 
			The rectangle in each panel represents the spatial and velocity resolutions. 
			\label{fig:PV_MN}}
\end{figure}

\begin{figure}
	\iffigure
	\vspace*{-10pt}
	\epsscale{1.0}
	\includegraphics[viewport = 0 0 100 700, scale = 0.58]{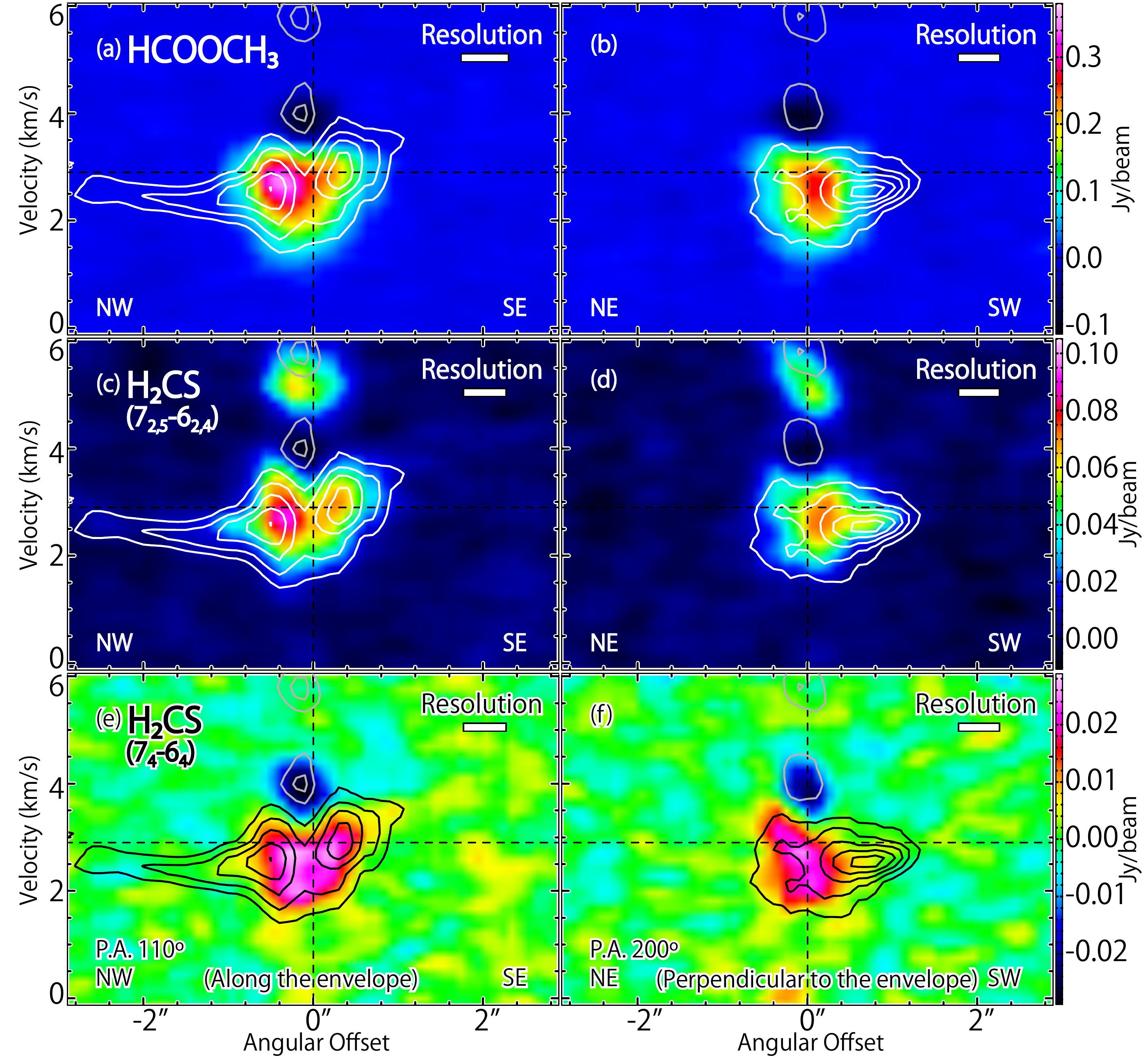}
	\fi
	\caption{The position-velocity diagrams of the \MF\ (\mfa) and two \TFA\ (\tfab; \tfae) lines. 
			The position axes are the same as those in the panels of P.As. of \PAenv\ and \PAoutflow\ in Figure \ref{fig:PV_OCS}. 
			The contours in each panel represent the PV diagram of \TFA\ (\tfaa), 
			which are as the same as those in Figure \ref{fig:PV_TFA}. 
			The rectangle in each panel represents the spatial and velocity resolutions. 
			\label{fig:PV_others}}
\end{figure}

Figures \ref{fig:PV_OCS} to \ref{fig:PV_others} show the position-velocity (PV) diagrams of OCS, \TFA, \MN, and \MF\ that 
allow us to investigate the velocity structure around the protostar. 
In the PV diagrams, the \iPC\ is confirmed toward the protostar. 
Along the PA of \PAenv, a slight velocity gradient can be seen in the OCS line (Figure \ref{fig:PV_OCS}), 
although it is heavily contaminated with the absorption feature toward the continuum peak position. 
The velocity in the northwestern side tends to be lower than that in the southeastern side. 
In this diagram, the position of the intensity peak has a slight offset from the continuum peak position. 
The PV diagrams along the lines with various PAs also show the absorption feature toward the continuum peak position. 
Moreover, the PV diagrams along the lines with the PAs of 140\degr, 170\degr, 200\degr, and 230\degr\ 
show two intensity peaks with an offset from the continuum peak position. 
For instance, the peak intensity ($\sim$0.6 Jy beam\inv) is about 1.5 times higher than the intensity toward the continuum peak position 
($\sim$0.4 Jy beam\inv) in the PV diagram along the line with the PA of 200\degr.
	
The PV diagram of the \TFA\ (\tfaa) line along the PA of \PAenv\ shows 
a slight velocity gradient similar to the OCS case (Figure \ref{fig:PV_TFA}). 
The absorption in the red-shifted component toward the continuum peak can be confirmed. 
There is another absorption feature with the velocity higher than 5 \kmps, 
which is likely to be a contamination by an unidentified line. 
As in the case of OCS, 
the \TFA\ line also shows two intensity peaks in the PV diagrams except for 
that along the PA of \PAoutflow\ (i.e. the direction perpendicular to the disk/envelope direction). 
The intensity dip toward the continuum peak position is clearer in \TFA\ than in OCS. 
In fact, the peak integrated intensity ($\sim$0.57 Jy beam\inv\ \kmps) is 
twice higher than the intensity toward the continuum peak (0.45 Jy beam\inv\ \kmps) 
in Figure \ref{fig:intprofile}. 
The peak intensities have asymmetry in the PV diagrams; 
the peak intensity at the southern side of the protostar is higher than at the northern side, 
except for the PV diagram along the PA of \PAenv. 
This may be caused by the asymmetric distribution of the gas. 

On the other hand, the PV diagrams of the \MN\ and \MF\ lines do not show such a clear velocity gradient, 
as shown in Figures \ref{fig:PV_MN} and \ref{fig:PV_others}(a, b), respectively. 
The absence of the apparent velocity gradient for these lines seems to originate from the contamination of the \iPC. 
The distributions of the \MN\ and \MF\ emission along the PA of \PAenv\ is as compact as 1\farcs5. 
Their maximum velocity shift from the systemic velocity (\aboutVsys) is 2 \kmps, 
which is comparable to that for OCS. 
The distributions for the \MN\ and \MF\ lines are so compact that 
the red-shifted part of their emission near the continuum peak is heavily obscured by the unresolved absorption feature of the \iPC. 
We also prepare the moment 1 maps (not shown in this article) 
of the \MN\ and \MF\ lines to closely inspect the velocity gradient in these lines. 
However, no significant features of the velocity gradient can be seen for these lines. 
It should be noted that the compact distributions of the \MN\ and \MF\ emission are not due to excitation effect, 
because the upper-state energies of these lines are comparable to or even lower than that of OCS (See Table \ref{tb:lines}). 

In Figure \ref{fig:PV_MN}, all the six PV diagrams of \MN\ show some asymmetry with respect to the continuum peak position. 
They are essentially similar to one another in the sizes of the absorption and the emission. 
This asymmetry in the intensities seems to be anti-correlated to that seen in the \TFA\ line 
for the PV diagrams along the PA of (140--200)\degr. 
Although the reason of this anticorrelation is puzzling at present, 
such a chemical differentiation could be a key to understand chemical processes occurring there. 

\subsection{Comparison of the Observed Molecular Distributions with the Source A Case} \label{sec:sourceA}
Since rotation motion around Source B is suggested by the OCS and \TFA\ emission, 
we investigate its kinematic structure in more detail. 
As for Source A, we successfully explained the kinematic structure of the \ire\ at a 100 au scale by a ballistic model \citep{Oya_15398, Oya_16293A}. 
Hence, we take the advantage of this knowledge in the analysis of the Source B data. 

In Source A, OCS traces the \ire, while the COMs highlight the \cb, as mentioned in Section \ref{sec:16293}. 
This situation is consistent with the chemical differentiation in Source B, 
where the OCS distribution is rather extended while the \MN\ and \MF\ distributions are compact around the protostar. 
If the OCS line traces the \ire, 
the velocity gradient shown in Figure \ref{fig:PV_OCS} can be interpreted 
by a combination of the infall and rotation motions in the envelope. 
The size of the \MN\ and \MF\ emission would be close to the size of the \cb, as in the case of Source A. 
These features will be examined by our kinematic model in Section \ref{sec:analysis}. 

On the other hand, 
the distribution of \TFA\ in Source B inside the \cb\ seems 
different from that in Source A. 
In Source A, 
high velocity-shift components tracing the disk component inside the \cb\ are observed. 
In contrast, such components are not apparently seen in Source B. 
The PV diagram shows a double-peaked structure having an intensity dip at the continuum peak position 
(Figure \ref{fig:PV_TFA}). 
This dip structure does not originate from the high dust opacity, 
because the emission, 
particularly at the blue-shifted one, 
is indeed observed toward the continuum peak position 
in other molecular lines: 
the PV diagrams of the \MN, \MF, and \TFA\ (\tfab; \tfae) lines show a compact emission in Figures \ref{fig:PV_MN} and \ref{fig:PV_others}. 
The dip is not due to the absorption by the foreground gas either, 
because the dip is seen at the velocity shift of 
$-1$ \kmps\ from the systemic velocity (\aboutVsys). 
Hence, the intensity dip means that \TFA\ would be deficient in the closest vicinity of the protostar. 
The two intensity peaks in the PV diagram just correspond to the position within which \MN\ and \MF\ appear (Figures \ref{fig:PV_others}a, b). 
Although the double-peaked structure is not clearly seen in OCS (Figure \ref{fig:PV_OCS}), 
the intensity peak of OCS seems to be shifted from the continuum peak position 
and almost coincides one of the intensity peaks of \TFA. 
Considering that \MN\ and \MF\ are concentrated around the \cb\ in Source A, 
it is most likely that the two intensity peaks in the PV diagram of \TFA\ represent 
the positions of the \cb\ in Source B. 

Figure \ref{fig:geometry}(a) shows a schematic illustration of the above configuration of the disk/envelope system. 
The structure of the outflow is consistent with the configuration, 
as described later (Section \ref{sec:outflow}). 

\section{Modelling} \label{sec:analysis}
\subsection{Infalling-Rotating Envelope Model} \label{sec:model_ire}
We investigate the velocity gradient observed in the OCS (\ocs) and \TFA\ (\tfaa) lines, 
using the \ire\ model by \citet{Oya_15398}. 
Since the velocity gradient is seen as the double-peaked structure 
in the PV diagram of \TFA\ (\tfaa) along the disk/envelope direction (PA \PAenv), as mentioned above, 
we evaluate the physical parameters by comparing the model with this PV diagram. 

\begin{figure}
	\iffigure
	\epsscale{1.0}
	\includegraphics[viewport = 0 0 100 700, scale = 0.4]{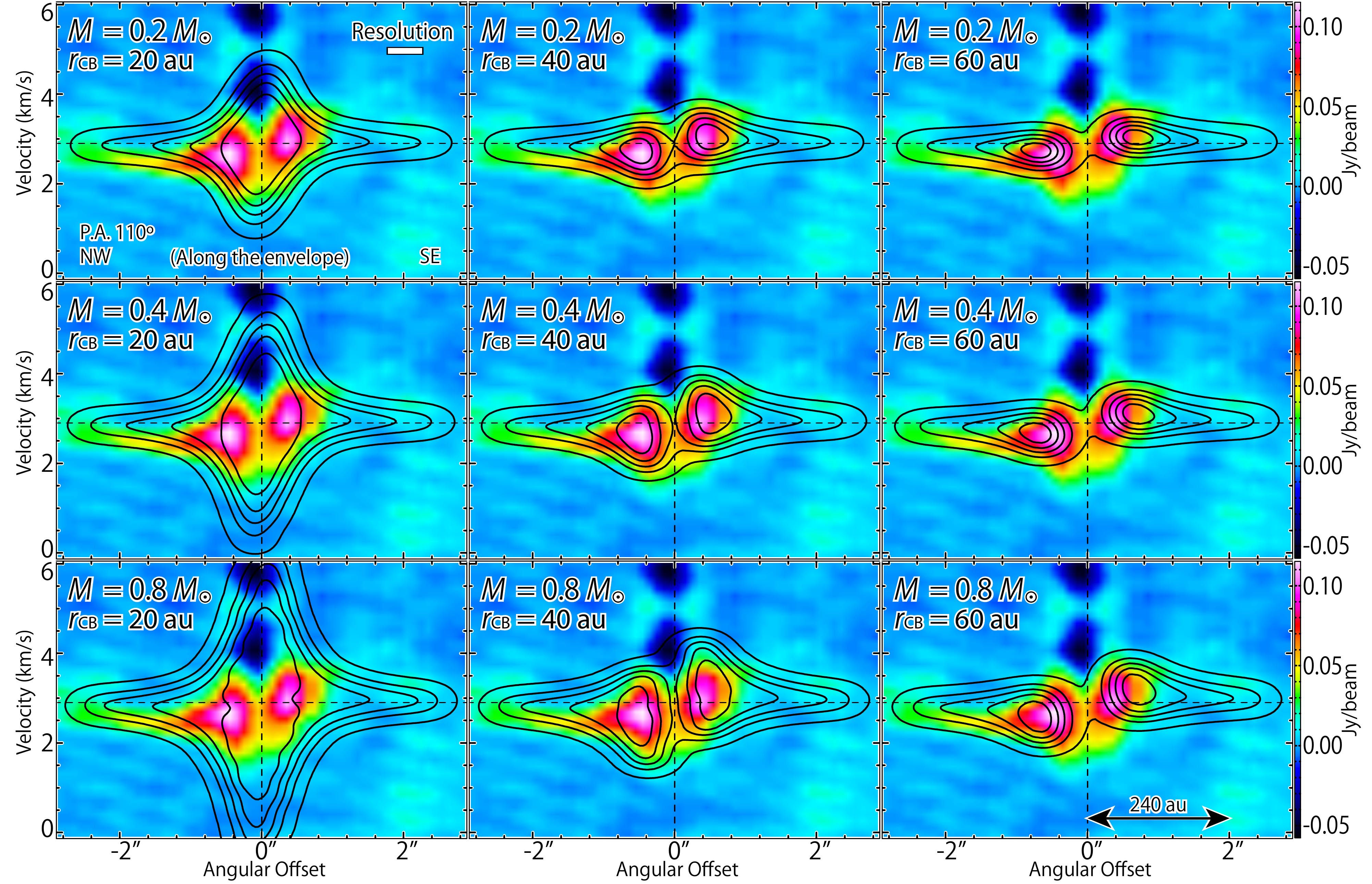}
	\fi
	\caption{The position-velocity diagrams of the \TFA\ (\tfaa; color) along the disk/envelope direction (PA \PAenv), 
			where the color maps are the same as the panel of a PA \PAenv\ in Figure \ref{fig:PV_TFA}. 
			The black contours represent the results of \ire\ models. 
			The physical parameters for the models are: 
			$M$ = 0.2, 0.4, and 0.8 \Msun; 
			\rcb\ = 20, 40, and 60 au; 
			and $i$ = \parI. 
			The contour levels are every 20\%\ from 5\%\ of each peak intensity. 
			The rectangle in the top left panel represents the spatial and velocity resolutions. 
			\label{fig:PV_H2CS-model_PAenv}}
\end{figure}

\begin{figure}
	\iffigure
	\epsscale{1.0}
	\includegraphics[viewport = 0 0 100 700, scale = 0.4]{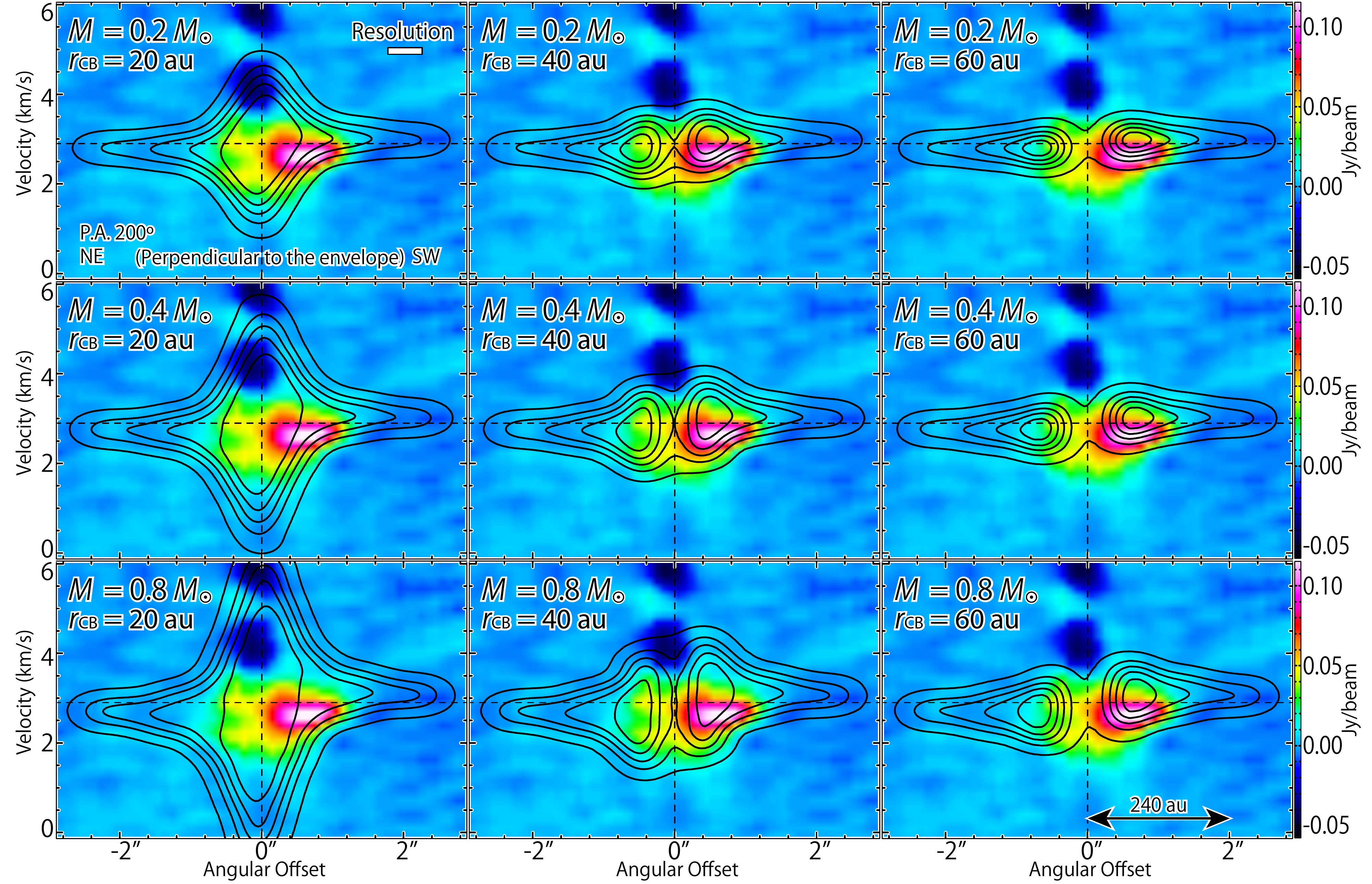}
	\fi
	\caption{The position-velocity diagrams of the \TFA\ (\tfaa; color) line 
			along the line perpendicular to the disk/envelope direction (PA \PAoutflow), 
			where the color maps are the same as the panel of a PA \PAoutflow\ in Figure \ref{fig:PV_TFA}. 
			The black contours represent the results of \ire\ models. 
			The physical parameters for the models are: 
			$M$ = 0.2, 0.4, and 0.8 \Msun; 
			\rcb\ = 20, 40, and 60 au; 
			and $i$ = \parI. 
			The contour levels are every 20\%\ from 5\%\ of each peak intensity. 
			The rectangle in the top left panel represents the spatial and velocity resolutions. 
			\label{fig:PV_H2CS-model_PAoutflow}}
\end{figure}

Figures \ref{fig:PV_H2CS-model_PAenv} and \ref{fig:PV_H2CS-model_PAoutflow} show examples of the \ire\ models 
superposed on the PV diagram of the \TFA\ (\tfaa) line along the disk/envelope direction 
and the direction perpendicular to it, respectively. 
We here adopt the systemic velocity of \parVsys\ in the model. 
In the \ire\ model, the main physical parameters are the protostellar mass ($M$) and the radius of the \cb\ (\rcb). 
To see how the PV diagrams of the \ire\ model depend on these parameters, 
we conduct the simulations for various sets of the parameters. 
We assume an envelope with a constant thickness of \parTau\ and the outer radius of \parRTFAau\ 
with the inclination angle of \parI\ (\incRem). 
We also assume the intrinsic line width of \parLW. 
The model image is convolved with the synthesized beam. 
Since we are interested in the kinematic structure around the protostar, 
the emissivity is simply assumed to be proportional to the \hydro\ density (\nhydro\ $\propto r^{-1.5}$) in the \ire, 
and no radiative transfer effect is considered in this model. 
Namely, we do not make fine tuning of the intensity distribution, 
but we rather just focus on the fundamental characteristics of the kinematic structure. 
Since the red-shifted components in these observations suffer from the absorption feature of the \iPC, 
we mainly consider the kinematic structure of the blue-shifted components. 

As shown in Figure \ref{fig:PV_H2CS-model_PAenv}, 
the velocity gradient along the disk/envelope direction seems to be reasonably explained by the rotating motion in the \ire\ model. 
Especially, the models with \rcb\ of 40 au well reproduce the observations except for the absorption feature. 
Although the model depends on the protostellar mass ($M$), 
it is not well constrained due to the contamination of the absorption feature. 
Hence, it is estimated to be around \parM\ within a factor of 2 
assuming the inclination angle of the disk/envelope system of \parI. 
On the other hand, the model result strongly depends on \rcb. 
In the models with \rcb\ of 60 au, the distance between the two intensity peaks is larger than that observed, 
because the positions of the two peaks in the model correspond to those of the \cb. 
In the models with \rcb\ of 20 au, the emission seems to be concentrated toward the protostar. 
This is because the emission at the \cb\ is not resolved with the synthesized beam in the model, 
and thus the observed double-peaked structure cannot be reproduced. 
Hence, \rcb\ is evaluated to be $(30-50)$ au. 
In the PV diagram along the line perpendicular to the disk/envelope direction (PA \PAoutflow; Figure \ref{fig:PV_H2CS-model_PAoutflow}), 
the models with \rcb\ of 40 au show a better agreement with the observations than those with \rcb\ of 20 or 60 au, 
although the observations show an asymmetric distribution of the emission. 

The inclination angle of \parI\ is fixed in the above analysis, 
where the northeastern side of the \desys\ faces to us. 
The velocity gradient cannot be explained with the inclination angle of 0\degr\ (completely face-on configuration) 
or a negative value, where the southwestern side of the \desys\ faces to us. 
Likewise, 
the velocity structure cannot be reproduced either with the inclination angle larger than 15\degr.

\begin{figure}
	\iffigure
	\epsscale{1.0}
	\includegraphics[viewport = 0 0 100 700, scale = 0.58]{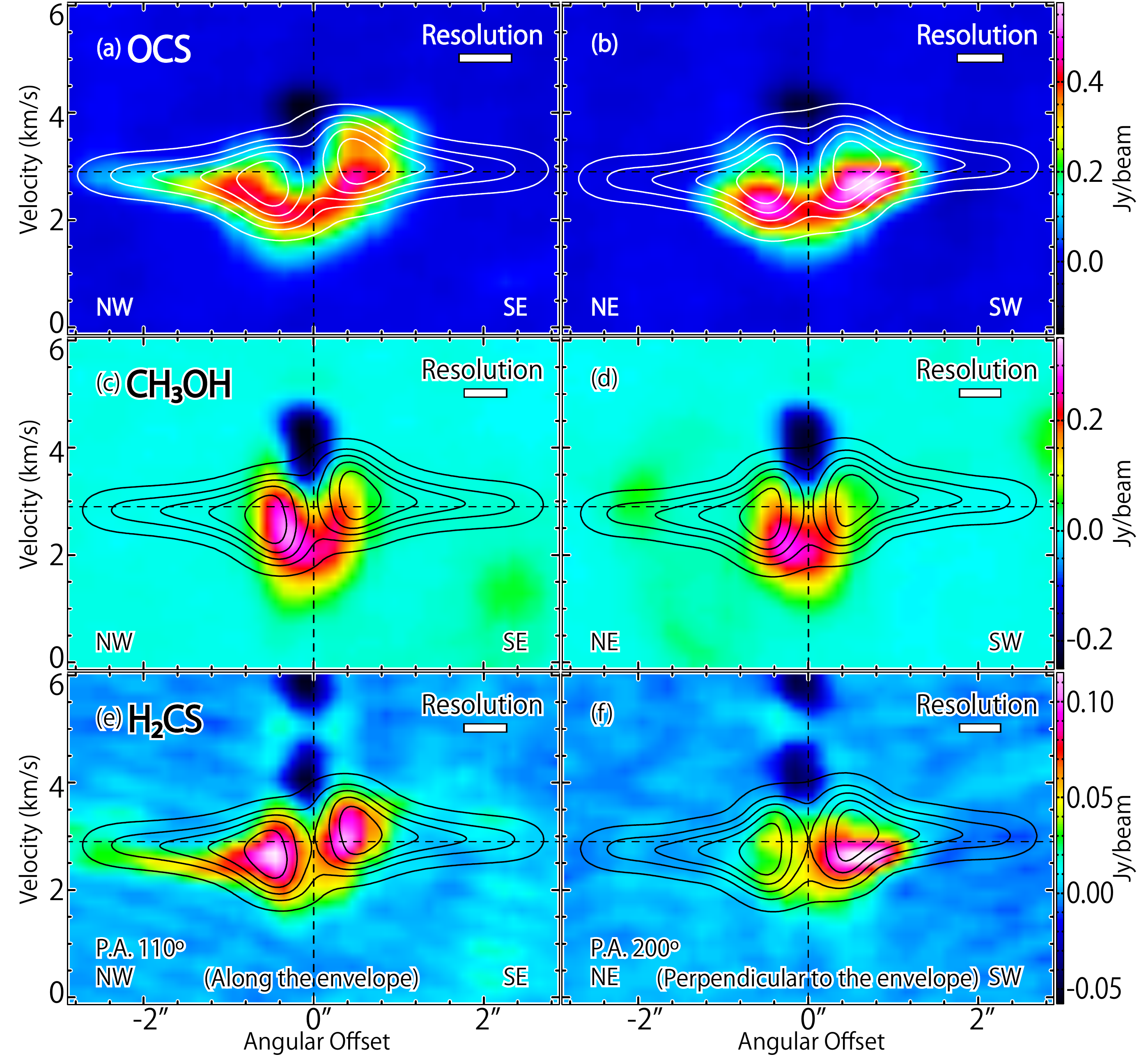}
	\fi
	\caption{The position-velocity diagrams of the OCS (\ocs; a, b), \MN\ (\mna; c, d), and \TFA\ (\tfaa; e, f) lines
			along the disk/envelope direction (PA \PAenv) and 
			the direction perpendicular to it (PA \PAoutflow), 
			where the color maps are the same as the panels (a, d) in Figures \ref{fig:PV_OCS}--\ref{fig:PV_TFA}.  
			The black and white contours represent the results of the \ire\ model, 
			where the physical parameters are $M$ = \parM, \rcb\ = \parRcbau, and $i$ = \parI. 
			The contour levels are every 20\%\ from 5\%\ of each peak intensity. 
			The rectangle in each panel represents the spatial and velocity resolutions. 
			\label{fig:PV_IREbest}}
\end{figure}

The comparison between the \ire\ model and the OCS (\ocs) and \MN\ (\mna) lines are also shown in Figure \ref{fig:PV_IREbest}. 
Here the model with $M$ of \parM\ and \rcb\ of \parRcbau\ is employed as a representative set of the parameters. 
Strictly speaking, the OCS (\ocs) line does not show a clear double-peaked feature unlike the \TFA\ (\tfaa) line. 
Nevertheless, the model seems to explain the basic velocity feature of the PV diagram of the OCS (\ocs) line, 
except for the absorption feature. 
On the other hand, 
the kinematic structure traced by the \MN\ (\mna) line cannot be explained by the \ire\ model 
obtained from the above analysis. 
The emitting region of the \MN\ (\mna) line seems to be concentrated around the \cb\ and/or inside it. 
This is consistent with the Source A case, 
where the \MN\ (\mnsourcea) line mainly highlights the \cb\ and/or inside it \citep{Oya_16293A}. 

\subsection{Origin of the inverse P-Cygni Profile} \label{sec:iPC}

Although the basic feature of the kinematic structure traced by the OCS and \TFA\ lines is reasonably explained 
by the \ire\ model (Figure \ref{fig:PV_IREbest}) as discussed above, 
there remains an important problem: 
{\it how can we interpret the \iPC?}
The \iPC\ means an infall motion along the line of sight toward the protostar. 
Since the disk/envelope system is nearly face-on, 
the outflow motion would also exist along the line of sight, 
as shown in Section \ref{sec:outflow}. 
This situation, namely the coexistence of the infall motion and the outflow motion along the line-of-sight in the vicinity of the protostar, 
is hardly possible, as far as we consider the thin disk structure such as Figure \ref{fig:geometry}(a). 
At least, 
a thick disk/envelope structure would be necessary so as 
that substantial amount of the infalling gas exists near the protostar. 
In fact, such an infall component perpendicular to the mid-plane 
near the launching point of the outflow 
is seen in numerical simulations of outflows \citep{Machida2013}. 
Even in this case, there remains the following difficulty. 
Since the velocity shift of the absorption feature from the systemic velocity (\aboutVsys) is at most 
2 \kmps\ (Figure \ref{fig:PV_MN}), 
the infalling gas at this velocity-shift has to be located at 180 au, 
assuming the free-fall with the protostellar mass of \parM. 
This distance is much larger than the apparent size of the COM emission ($\sim$50 au) around the protostar, 
which accompanies the \iPC. 

Recently, a hint to solve this problem is found in the other protostellar source L1527. 
On the basis of the high-resolution molecular line observations with ALMA, 
\citet{Sakai_1527_highres} reported that the thickness of the disk/envelope system is broadened at the \cb\ and inward of it 
due to the stagnation of the accreting gas. 
A part of the stagnant gas at the \cb\ once moves toward the out-of-plane direction, and falls toward the protostar, 
if its angular momentum is extracted by some mechanisms (i.e. disk winds and/or low-velocity outflows \citep[c.f.,][]{Alves_BHB07-11}). 
If this situation is also the case for Source B, 
such an infalling gas may cause the \iPC. 
This situation is schematically illustrated in Figure \ref{fig:geometry}(b). 

\begin{figure}
	\iffigure
	\epsscale{1.0}
	\includegraphics[viewport = 0 0 100 730, scale = 0.58]{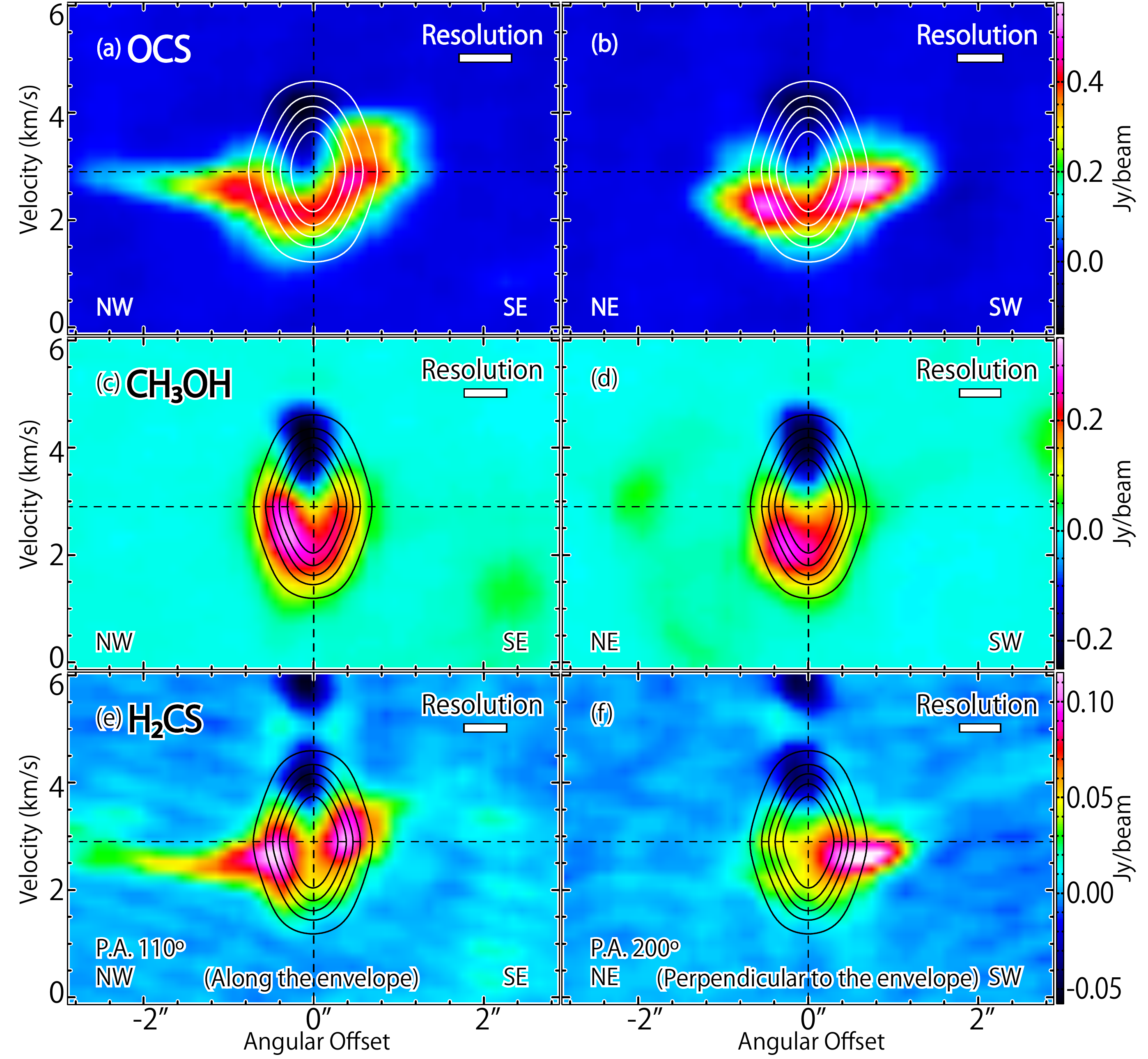}
	\fi
	\caption{The position-velocity diagrams of the OCS (\ocs; a, b), \MN\ (\mna; c, d), and \TFA\ (\tfaa; e, f) lines
			along the disk/envelope direction (PA \PAenv) and 
			the direction perpendicular to it (PA \PAoutflow), 
			where the color maps are the same as those in Figures \ref{fig:PV_IREbest}.  
			The black and white contours represent the results of the model for the infall motion from the \cb. 
			The physical parameters are $M$ = \parM, \rcb\ = \parRcbau, and $i$ = \parI. 
			The molecular line emission is assumed to be in the region with the distance from the protostar of (\parRfall--\parRcb) au. 
			The contour levels are every 20\%\ from 5\%\ of each peak intensity. 
			The rectangle in each panel represents the spatial and velocity resolutions. 
			\label{fig:PV_freefall}}
\end{figure}

Hence, we incorporate the free-fall motion in the vicinity of the protostar in the model. 
We approximate the gas distribution around the protostar as a spherical clump with a radius of the \cb\ for simplicity. 
This model simulates the hypothetical situation that the infalling gas is once stagnated at the distance of \parRcbau\ (\rcb) from the protostar to make a spherical clump, 
and then it falls to the protostar by the gravity without the initial radial speed. 
Here, the `spherical clump' mimics the gas stagnation, 
although the gas distribution is not spherical in reality. 
Figure \ref{fig:PV_freefall} shows the results. 
This very simplified picture can explain the kinematic structure of \MN\ at a \parRcbau\ scale around the protostar, 
including the velocity of the absorption in the \iPC. 
The kinematic structures of OCS and \TFA, on the other hand, 
seem to be explained by the combination of the \ire\ model for the extended components 
and this free-fall picture for the \iPC. 
In this free-fall picture, the absorption gas for the \iPC\ having the infall velocity of 2 \kmps\ is located at 33 au from the protostar, 
which is smaller than the size of their apparent distribution. 
However, it should be stressed that this model is just a simplified one representing 
the above hypothetical physical picture. 
We need to resolve the structure of the vicinity of the protostar to verify its validity. 

\section{Outflow} \label{sec:outflow} 
\begin{figure}
	\iffigure
	\epsscale{1.0}
	\includegraphics[viewport = 0 0 100 500, scale = 0.75]{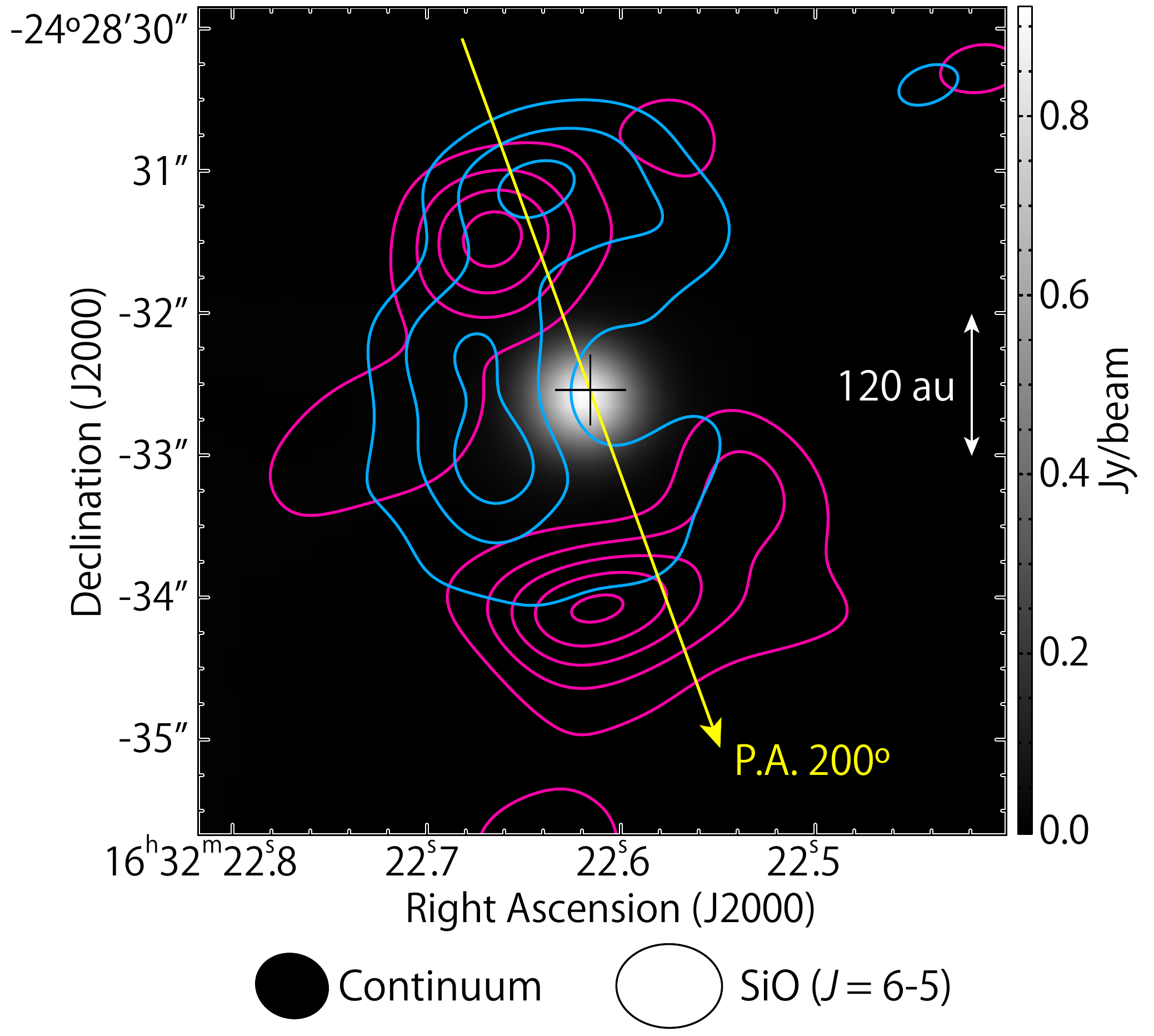}
	\fi
	\caption{The integrated intensity maps of high velocity components of the SiO (\sio) line. 
			The red contours represent the integrated intensity map of SiO with the velocity range from 3.8 to 5.8 \kmps, 
			while the blue contours with the velocity range from $-0.2$ to 1.7 \kmps. 
			The contour levels are every 10$\sigma$, where the rms noise level is 20 mJy beam\inv\ \kmps. 
			The gray scale map represent the continuum map, and the intensity peak position in Source B is shown by the black cross. 
			The synthesized beams are depicted {\bf below the map.} 
			The PA of \PAoutflow\ perpendicular to the disk/envelope direction is shown (see Section \ref{sec:kin_rot}). 
			\label{fig:highV_SiO}}
	\vspace*{-20pt}
\end{figure}

We have also analyzed the SiO (\sio) data. 
Figure \ref{fig:highV_SiO} shows the integrated intensity maps of the high velocity-shift components traced by SiO. 
This molecule is often employed as a shock tracer \citep[e.g.,][]{Mikami_L1157-SiO, Bachiller_shockL1157}. 
Both the blue-shifted and red-shifted components show a shell-like feature surrounding the continuum peak, 
which are spatially overlapped with each other even at a 300 au scale. 
\citet{Kristensen_16293outflow} have suggested by using the CO ($J=6-5$) ALMA observation 
that there is an interaction of the outflow from Source A with Source B, resulting in shocks there. 
In addition, 
\citet{Girart_16293outflow} have also detected the SiO ($J=8-7$) line around Source B, 
and have suggested 
that the SiO morphology is consistent with an 
interaction of the outflow from Source A with Source B. 
Nevertheless, the authors did not investigate the kinematics. 
Although this possibility cannot be ruled out, 
these SiO components can also be attributed to the outflow from Source B with a nearly pole-on geometry; 
namely, its kinematic structure is consistent with the situation that 
we are looking at the outflow cavities as a shell-like feature. 
We favor this possibility, 
because the CO outflow from Source A is blue-shifted while the SiO emission has the red-shift component. 
Moreover, the interaction cannot be seen clearly in the other lines. 
For instance, no enhancement of \MN\ is seen in the southeastern side (Figure \ref{fig:PV_IREbest}c). 
Furthermore, the SiO emission can be seen in the back side of Source B 
with respect to the direction to Source A. 

\begin{figure}
	\iffigure
	\epsscale{0.9}
	\includegraphics[viewport = 0 0 100 700, scale = 0.4]{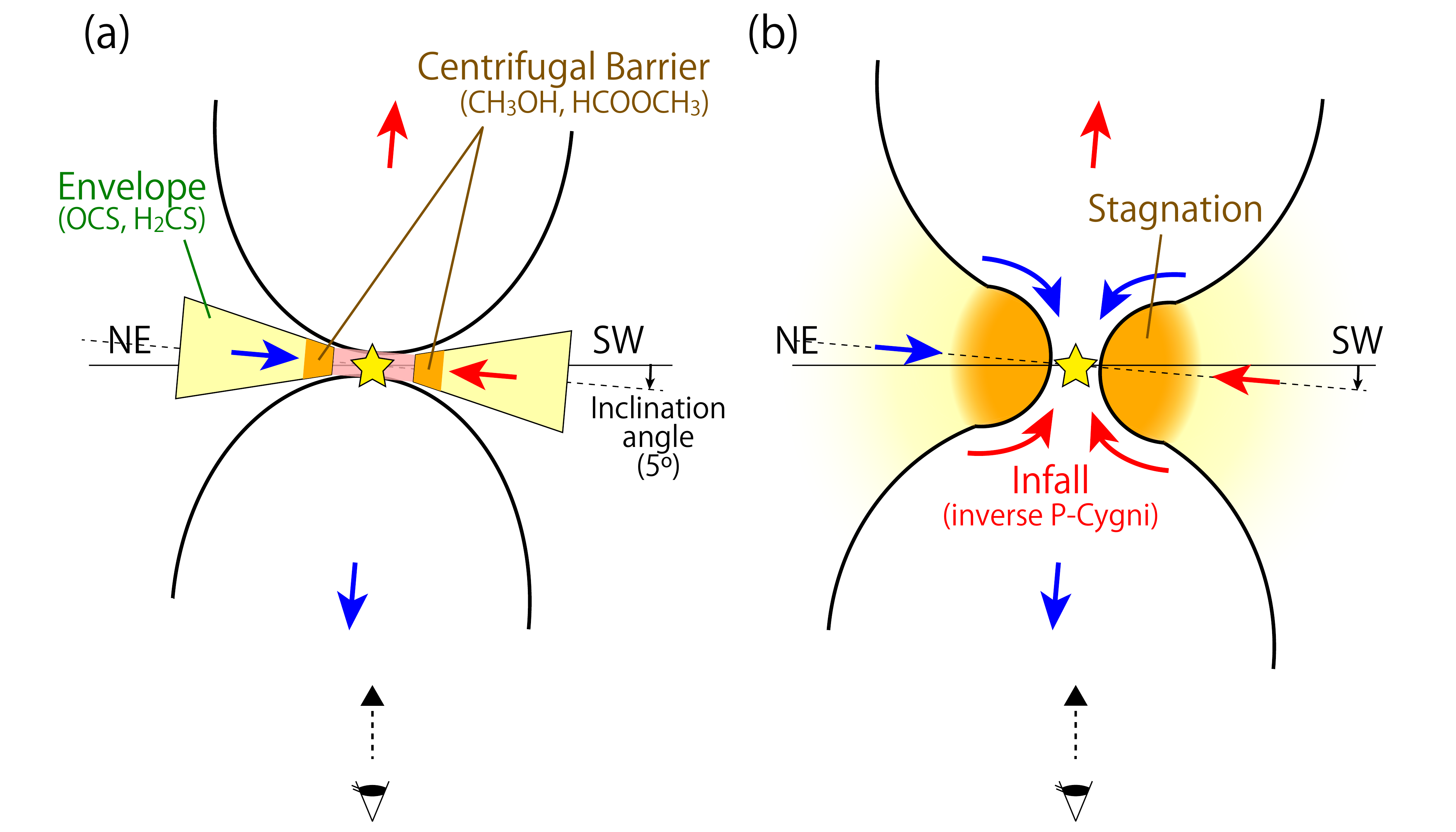}
	\fi
	\caption{The schematic illustration of the disk/envelope system in Source B. 
			The disk/envelope system has a thin structure in panel (a), 
			while it has an infalling gas around the protostar in panel (b). 
			\label{fig:geometry}} 
\end{figure}

When we look at the two components of SiO closely (Figure \ref{fig:highV_SiO}), 
the blue-shifted components are slightly extended toward the northeastern direction from the continuum peak. 
On the other hand, the red-shifted emission is stronger than the blue-shifted one in the southwestern side of the continuum peak. 
Hence, the outflow axis might be inclined slightly (positive inclination angle), 
as illustrated in Figure \ref{fig:geometry}. 
Thus, the blue-shifted outflow lobe seems to be overlapped on the protostar along the line of sight. 
This interpretation is natural, 
because the \desys\ has a face-on geometry. 
The positive inclination angle is consistent with that suggested by the analysis of the \desys. 

\begin{figure}
	\iffigure
	\epsscale{1.0}
	\includegraphics[viewport = 0 -100 100 2500, scale = 0.11]{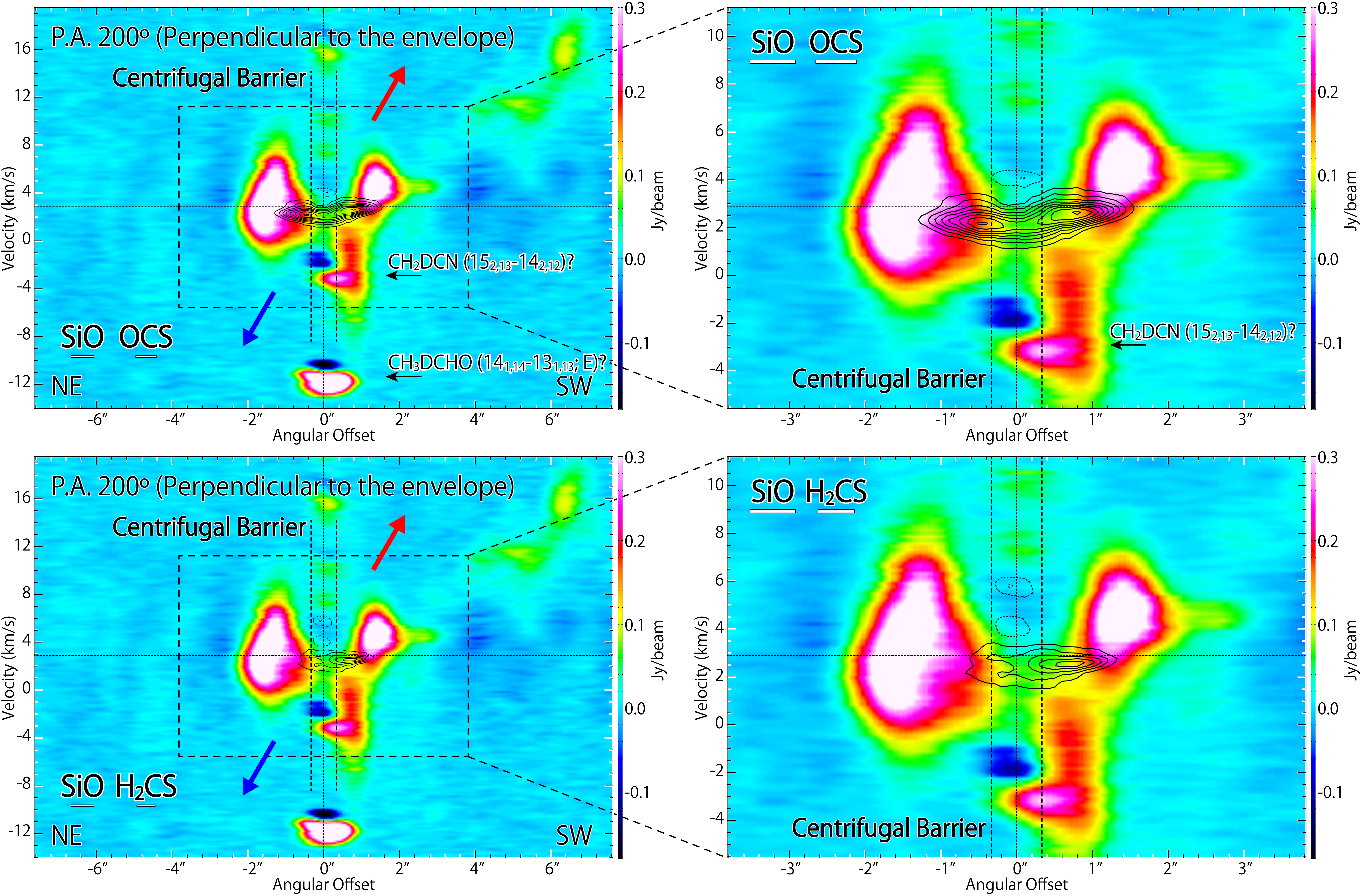}
	\vspace*{-20pt}
	\fi
	\caption{The position-velocity diagram of the SiO (\sio; color), OCS (\ocs; black contours in the top {\bf panels}), 
			and \TFA\ (\tfaa; black contours in the bottom {\bf panels}) 
			along the line centered at the continuum peak with the PA of \PAoutflow. 
			{\bf The panels in the right column are the blown-ups of those in the left column.} 
			The absorption features at the velocity of 2 and 10 \kmps\ in the color map seem to be 
			the contamination by other molecular lines with the \iPC. 
			The maps of the OCS and \TFA\ lines are shown only for the velocity range from 0 to 6 \kmps. 
			The red and blue arrows represent the outflow directions judged from Figure \ref{fig:highV_SiO}. 
			The rectangles in each panel represent the spatial and velocity resolutions. 
			\label{fig:PV_SiO-OCS_PAoutflow}} 
\end{figure}

In the OCS, \MN, \MF, and \TFA\ lines, the \iPC\ is seen toward the protostar. 
As discussed in Section \ref{sec:iPC}, 
the picture shown in Figure \ref{fig:geometry}(b) allows the coexistence of the blue-shifted outflow lobe and the infalling gas toward the protostar. 
In this case, the outflow responsible for the above SiO distribution would be not launched directly from the protostar, 
but could be from the inner part of the disk/envelope system possibly around the \cb. 
In fact, there is a hint for such a situation in the kinematic structure of the SiO emission. 
The SiO emission shows the bipolar outflow lobes along the PA of \PAoutflow, as shown in Figure \ref{fig:PV_SiO-OCS_PAoutflow}. 
In this figure, we note that there are the two absorption features toward the continuum peak position, 
which are likely due to the contamination by other molecular lines with the \iPC. 
The blue-shifted and red-shifted lobes are seen in both the northeastern and southwestern sides of the continuum peak, 
which is consistent with the configuration shown in Figure \ref{fig:geometry}(b). 
If the outflow is accelerated as it propagates away from the protostar \citep[e.g.][]{Lee_outflow}, 
the systemic velocity (\aboutVsys) component 
may approximately regarded as 
the outflow component at its launching point. 
At the systemic velocity, the SiO emission is seen at the position with an offset of $\sim$0\farcs5 from the continuum peak, 
but not toward the protostar. 
The radial size of the SiO dip is larger than those of the OCS and \TFA\ emission in 
Figure \ref{fig:PV_SiO-OCS_PAoutflow}. 
Hence, this feature is not likely to be due to the optical depth effect toward the continuum peak. 
This implies that the launching point of the outflow has a radial offset from the protostar (Figure \ref{fig:geometry}b). 
Moreover, the SiO emission at the systemic velocity seems to appear near to the intensity peak of 
the OCS and \TFA\ emission. 
This feature may suggest that the launching point of the outflow could be around the \cb\ traced by OCS and \TFA. 
A similar situation is recently reported for BHB07-11 \citep{Alves_BHB07-11}. 

Large-scale outflows blowing out from this binary system have extensively been studied 
\citep[e.g.][]{Mizuno_largeoutflow, Hirano_largeoutflow}. 
Their directions are different from that found in this study. 
This contradiction may be due to the complexity of the binary system. 
The dynamical timescale of the outflow lobes at a larger scale $(10^4 - 10^5)$ au 
is estimated to be $(10^4 - 10^5)$ years \citep[]{Mizuno_largeoutflow, Hirano_largeoutflow}. 
On the other hand, it is reported that Source A and Source B are rotating around each other with the period of 
$\sim 2 \times 10^4$ years \citep[]{Bottinelli_Mass}. 
Thus the directions of the two outflow can be modulated in a complex way. 
Hence, the small-scale outflow structure in the vicinity of a protostar has to be studied 
in order to explore the relation between the outflow and the disk/envelope system. 

It should be noted that a small-scale outflow structure of \iras\ is investigated by using the CO lines with ALMA 
\citep{Kristensen_16293outflow, Loinard_16293outflow, Girart_16293outflow}. 
However, the outflow morphology seen in the CO lines 
in this source is very complicated. 
The northwest-southeast outflow is seen in Source A, 
while the outflow from Source B is unclear in these observations. 
In contrast, our kinematic analysis suggests 
a possibility that the SiO emission traces the outflow launched from Source B. 

\section{Gas Kinetic Temperature} \label{sec:Tkin}
For Source A, \citet{Oya_16293A} evaluated the gas kinetic temperature from the intensities of the two lines (\tfaa; \tfab) of \TFA. 
The lines of \TFA\ with different \Ka\ can be used as a good tracer of the gas kinetic temperature. 
As a result, it is found that the gas kinetic temperature around Source A once rises from the \ire\ to the \cb, 
and then drops in the disk component inside the \cb. 

\begin{figure}
	\iffigure
	\epsscale{1.0}
	\includegraphics[viewport = 0 0 100 300, scale = 1.04]{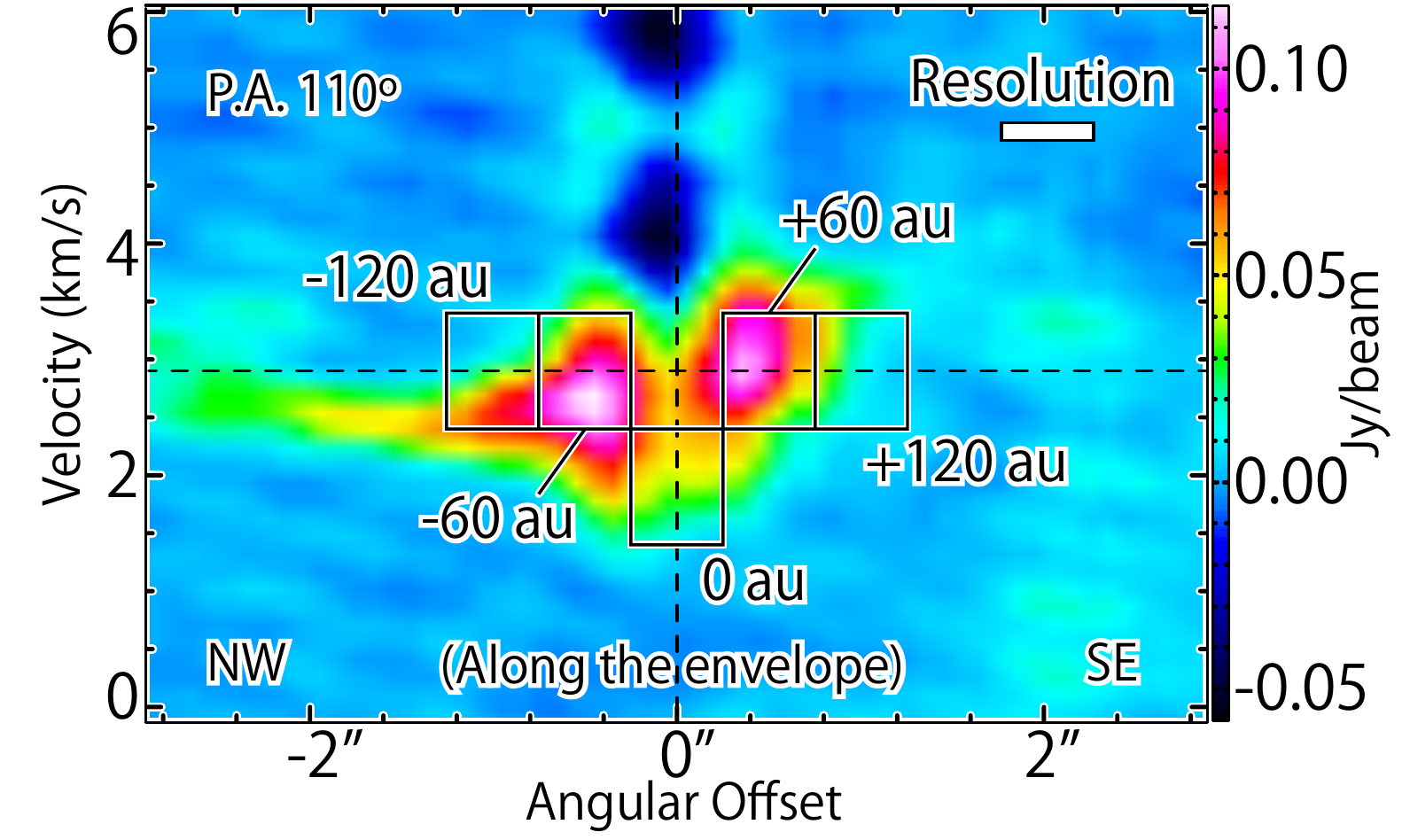}
	\fi
	\caption{The position-velocity diagram of the \TFA\ (\tfaa) line along the disk/envelope direction (PA \PAenv). 
			The vertical edges of the black rectangles represent the velocity ranges which are integrated to derive the gas kinetic temperature in Section \ref{sec:Tkin}. 
			The horizontal edges of them are centered at the positions at the distance of 0, $\pm$\offsetCB, and $\pm$\offsetEnv\ au 
			from the continuum peak position along the disk/envelope direction (PA \PAenv), 
			where the gas kinetic temperature is derived. 
			The lengths of the horizontal edges of them correspond to the angular resolution for the \TFA\ (\tfaa) line. 
			The rectangle represents the spatial and velocity resolutions. 
			\label{fig:PV_H2CS_Tkin}}
\end{figure}

Here, we also evaluate the gas kinetic temperature around Source B from the \TFA\ line intensities. 
Figures \ref{fig:PV_others}(c--f) show the PV diagrams of the high excitation lines of \TFA\ (\tfab; \tfae) 
on which those of the \TFA\ line (\tfaa) is superposed. 
The distributions of the higher excitation lines of \TFA\ seem to be 
more concentrated to the continuum peak position 
than the \TFA\ line (\tfaa). 
Thus the intensity ratio of a higher excitation line relative to the lower excitation line 
becomes higher toward the protostar. 
This implies that the gas kinetic temperature is raised as approaching to the continuum peak position. 
The gas kinetic temperature is evaluated 
by using the non-LTE code RADEX \citep{vanderTak_radex}, as shown in Table \ref{tb:Tkin}. 
Temperatures are calculated for the positions with the radii of 0, \offsetCB, and \offsetEnv\ au 
from the continuum peak along the disk/envelope direction (PA \PAenv). 
For this calculation, we prepare the integrated intensity maps of the three \TFA\ lines (\tfaa; \tfab; \tfae) with the velocity width of 1 \kmps. 
The velocity-shift range from $-1.5$ to $-0.5$ \kmps\ is taken at the continuum peak position 
in order to extract the high velocity-shift component excluding the absorption effect. 
The \TFA\ (\tfae) line emission is detected with this velocity range (Figure \ref{fig:PV_others}). 
On the other hand, the velocity-shift range from $-0.5$ to $+0.5$ \kmps\ is taken for the positions at the distance of 
\offsetCB\ and \offsetEnv\ au 
from the continuum peak position. 
The above velocity-ranges and the positions are shown in the PV diagram of \TFA\ (\tfaa) (Figure \ref{fig:PV_H2CS_Tkin}). 
It should be noted that the \TFA\ (\tfaa) line may possibly be optically thick toward the continuum peak. 
Hence, the intensity of the \TFA\ (\tfaa) line may be attenuated, 
which would overestimate the gas kinetic temperature. 
Thus the values derived from the \tfab\ and \tfae\ lines (the second row in Table \ref{tb:Tkin}) 
may be more reliable than those derived from the \tfaa\ and \tfab\ lines (the first row). 
The gas kinetic temperature is as high as $\sim$100 K within the radius of \offsetCB\ au. 
Moreover, the temperature tends to be higher as getting close to the protostar probably due to the thermal heating by the protostar. 
In Source B, we do not find any particular enhancement of the gas kinetic temperature at the \cb\ in contrast to the Source A case, 
probably because of the limited resolution of the observations. 
Higher angular resolution observations are required for further investigation 
on the small-scale temperature structure of this source.

\section{Summary} \label{sec:summary}
We analyzed the OCS, \MN, \MF, \TFA, and SiO data observed toward \iras\ Source B at a sub-arcsecond resolution with ALMA.
Major findings are as follows: 
	
\noindent(1) 
The chemical differentiation observed for the above molecules is similar to that found in Source A; 
OCS and \TFA\ have more extended distributions than \MN\ and \MF. 

\noindent(2)
Although the disk/envelope system of Source B has a nearly face-on configuration, 
a modest 
rotation feature is implied in the extended component of the \TFA\ and OCS lines. 
It is reasonably interpreted in terms of the \ire\ model assuming the ballistic motion. 
On the other hand, the \MN\ and \MF\ lines do not show a rotation feature. 
It might be due to the serious contamination of the unresolved absorption feature of the \iPC. 

\noindent(3) 
The bipolar outflow lobes near the protostar are likely detected in the SiO emission, 
although a possibility of the impact of the Source A outflow on Source B cannot be ruled out 
as the origin of the SiO emission. 
The blue and red lobes are largely overlapped, 
and hence, this feature is consistent with 
a nearly pole-on geometry 
of the \desys\ of this source. 

\noindent(4) 
The molecular lines, except for \MF, show absorption features toward the protostar. 
It is interpreted as the \iPC\ by the infalling gas along the line of sight, as previously reported. 
The infall motion is reasonably explained by the free-fall motion of the gas stagnated around the \cb, 
although it is a tentative and quite simplified interpretation. 

\noindent(5) 
The coexistence of the outflow and the infall motion toward the protostar 
may suggest that the current launching point of the outflow responsible for the SiO emission likely has an offset from the protostar. 

\noindent(6) 
The gas kinetic temperature is evaluated by using the \TFA\ lines. 
The gas kinetic temperature is found to be as high as $\sim$100 K, 
which is consistent to the hot corino character of \iras\ Source B. 
The temperature is systematically higher toward the continuum peak position than toward the outer positions. 
No enhancement of the temperature is seen around the \cb\ in contrast to the Source A case.

\acknowledgments
We thank the anonymous reviewer for valuable comments. 
This paper makes use of the ALMA data set ADS/JAO.ALMA\#2012.1.00712.S and \#2015.1.01060.S. 
ALMA is a partnership of the ESO (representing its member states), 
the NSF (USA) and NINS (Japan), together with the NRC (Canada) and the NSC and ASIAA (Taiwan), 
in cooperation with the Republic of Chile. 
The Joint ALMA Observatory is operated by the ESO, the AUI/NRAO and the NAOJ. 
The authors are grateful to the ALMA staff for their excellent support. 
Y.O. acknowledges the JSPS fellowship. 
This study is supported by Grant-in-Aid from the Ministry of Education, Culture, Sports, Science, and Technologies of Japan (21224002, 25400223, 25108005, and 15J01610). 
N.S. and S.Y. acknowledge financial support by JSPS and MAEE under the Japan--France integrated action program (SAKURA: 25765VC).
Y.O., N.S, Y.W., and S.Y. also acknowledge financial support by JSPS and MAEE under the Japan--France integrated action program.  
C.C. and B.L. acknowledge the financial support by CNRS under the France--Japan action program. 
C.F. acknowledges the support from the Italian Ministry of Education, Universities and Research, project SIR (RBSI14ZRHR).


\clearpage
\begin{landscape}
\begin{table}
	\begin{center}
	\caption{Parameters of the Observed Lines \label{tb:lines}}
	\vspace*{10pt} 
	\begin{tabular}{llcccc}
		\hline
		Molecule & Transition & Frequency (GHz) & $E_{\rm u}$ (K) & $S\mu^2$ (Debye$^2$)\tablenotemark{\mathrm a} & Synthesized Beam \\ \hline
		OCS\tablenotemark{\mathrm b, c} & \ocs\ & 231.0609934 & 111 & 9.72 & $0\farcs65 \times 0\farcs52$ (PA 84\fdegr49) \\ 
		CH$_3$OH\tablenotemark{\mathrm b, c} & \mna\ & 239.746253 & 49 & 3.89 & $0\farcs53 \times 0\farcs46$ (PA 71\fdegr45) \\ 
		HCOOCH$_3$\tablenotemark{\mathrm c, d} & \mfa\ & 250.25837 & 134 & 51.1 & $0\farcs61 \times 0\farcs46$ (PA 79\fdegr40) \\ 
		H$_2$CS\tablenotemark{\mathrm b, c} & \tfaa\ & 240.2668724 & 46 & 19.0 & $0\farcs53 \times 0\farcs46$ (PA 73\fdegr50) \\ 
		H$_2$CS\tablenotemark{\mathrm b, c} & \tfab\ & 240.5490662 & 99 & 17.5 & $0\farcs53 \times 0\farcs46$ (PA 73\fdegr67) \\ 
		H$_2$CS\tablenotemark{\mathrm b, c} & \tfad \tablenotemark{\mathrm e} & 240.3321897 & 257 &  12.8, 12.8 & $0\farcs53 \times 0\farcs46$ (PA 73\fdegr58) \\ 
		SiO\tablenotemark{\mathrm b, f} & \sio\ & 260.5180090 & 43.8 & 57.6 & $0\farcs74 \times 0\farcs59$ (PA 89\fdegr09) \\ 
		\hline
	\end{tabular}
	\tablenotetext{a}{Nuclear spin degeneracy is not included. } 
	\tablenotetext{b}{Taken from CDMS \citep{Muller_CDMS}. } 
	\tablenotetext{c}{The ALMA Cycle 1 archival data. The self-calibration is employed. } 
	\tablenotetext{d}{Taken from JPL \citep{Pickett_JPL}. } 
	\tablenotetext{e}{The rest frequency of these two lines are identical. They are denoted as \tfae\ in the text. } 
	\tablenotetext{f}{The ALMA Cycle 3 data. The self-calibration is not employed. } 
	\end{center}
\end{table}
\end{landscape}

\clearpage
\begin{landscape}
\begin{table}
	\begin{center}
	\caption{Gas Kinetic Temperature Derived from the \TFA\ Lines\tablenotemark{\mathrm a} 
			\label{tb:Tkin}}
	\begin{tabular}{c|ccccc}
	\hline 
	 & \multicolumn{5}{c}{Offset from the protostar\tablenotemark{\mathrm b}} \\ 
	Transitions & $-$\offsetEnv \tablenotemark{\mathrm c} & $-$\offsetCB \tablenotemark{\mathrm c} & 0 au\tablenotemark{\mathrm d} & $+$\offsetCB \tablenotemark{\mathrm c} & $+$\offsetEnv \tablenotemark{\mathrm c} \\ \hline 
	\tfab\ / \tfaa\tablenotemark{\mathrm e} & $55-96$ & $>338$ & $>400$ & $90-115$ & $48-167$ \\ 
	\tfae / \tfab\tablenotemark{\mathrm f} & $>113$ & $67-103$ & $88-164$ & $70-125$ & $-$ \\ 
	\hline 
	\end{tabular}
	\tablenotetext{a}{In K. 
					The gas kinetic temperatures are derived by using the RADEX code \citep{vanderTak_radex}. 
					The assumed ranges for the H$_2$ density and the column density of \TFA\ are from $10^7$ to $3\times10^9$ cm$^{-3}$ 
					and from $10^{13}$ to $10^{15}$ cm$^{-2}$, respectively. 
					The quoted errors represent 1$\sigma$. 
					$\sigma$ is estimated from the statistical error and does not contain the calibration error, 
					because it will be almost canceled in the intensity ratios. 
					Since the \tfaa\ line may be optically thick near the protostar, 
					the values in the second row are more reliable than those in the first row. 
					} 
	\tablenotetext{b}{The positive and negative values represent the offset from the continuum peak position 
					toward the southeastern and northwestern direction, respectively. } 
	\tablenotetext{c}{The velocity-shift range for the integrated intensities is from $-0.5$ to $+0.5$ \kmps.}
	\tablenotetext{d}{The velocity-shift range for the integrated intensities is from $-1.5$ to $-0.5$ \kmps.}
	\tablenotetext{e}{The gas kinetic temperature derived from the intensity ratio of the \TFA\ (\tfaa; \tfab) lines. } 
	\tablenotetext{f}{The gas kinetic temperature derived from the intensity ratio of the \TFA\ (\tfab; \tfae) lines. } 
	\end{center}
\end{table}
\end{landscape}

\clearpage

\clearpage

\clearpage

\clearpage

\clearpage

\clearpage

\clearpage

\clearpage

\clearpage

\clearpage

\clearpage

\clearpage

\clearpage

\clearpage

\clearpage

\clearpage

\clearpage

\clearpage

\clearpage

\clearpage

\end{document}